\documentclass[10pt,letterpaper]{article}
\usepackage[top=0.85in,left=2.75in,footskip=0.75in]{geometry}

\usepackage{amsmath,amssymb}

\usepackage[normalem]{ulem}

\usepackage{changepage}

\usepackage[utf8x]{inputenc}

\usepackage{textcomp,marvosym}

\usepackage{cite}

\usepackage{nameref,hyperref}

\usepackage{microtype}
\DisableLigatures[f]{encoding = *, family = * }

\usepackage[table]{xcolor}

\usepackage{array}

\newcolumntype{+}{!{\vrule width 2pt}}

\newlength\savedwidth

\raggedright
\usepackage[aboveskip=1pt,labelfont=bf,labelsep=period,justification=raggedright,singlelinecheck=off]{caption}

\makeatletter
\renewcommand{\@biblabel}[1]{\quad#1.}
\makeatother

\date{}

\usepackage{lastpage,fancyhdr,graphicx}
\usepackage{epstopdf}
\fancyheadoffset[L]{2.25in}
\fancyfootoffset[L]{2.25in}
\begin{document}
\vspace*{0.2in}

\begin{flushleft}
{\Large
\textbf\newline{Small field models with gravitational wave signature supported by CMB data} 
}
\newline
\\
Ira Wolfson\textsuperscript{1}
\; Ramy Brustein\textsuperscript{1}
\\
\bigskip
\textbf{1} Department of physics, Ben-Gurion University of the Negev, 8410500 Beer-Sheva, Israel
\bigskip

%
%





* irawolf@post.bgu.ac.il

\end{flushleft}
\section*{Abstract}
We study scale dependence of the cosmic microwave background (CMB) power spectrum in a class of small, single-field models of inflation which lead to a high value of the tensor to scalar ratio. The inflaton potentials that we consider are degree 5 polynomials, for which we precisely calculate the power spectrum, and extract the cosmological parameters: the scalar index $n_s$, the running of the scalar index $n_{\text{\tiny{run}}}$  and the tensor to scalar ratio $r$.  We find that for non-vanishing $n_{\text{\tiny{run}}}$ and for $r$ as small as $r=0.001$, the precisely calculated values of $n_s$ and $n_{run}$ deviate significantly from what the standard analytic treatment predicts. We study in detail, and discuss the probable reasons for such deviations. As such, all previously considered models (of this kind) are based upon inaccurate assumptions. We scan the possible values of potential parameters for which the cosmological parameters are within the allowed range by observations.  The 5 parameter class is able to reproduce all of the allowed values of $n_s$ and $n_{\text{\tiny{run}}}$ for values of $r$ that are as high as 0.001. Subsequently this study at once refutes previous such models built using the analytical Stewart-Lyth term, and revives the small field brand, by building models that do yield an appreciable $r$ while conforming to known CMB observables.


\section*{Introduction}
Recent years have shown an increase in cosmological observational data, largely due to the Planck mission \cite{Ade:2015xua}, and the searches for primordial gravitational waves (GW) signal in the cosmic microwave background (CMB) by terrestrial experiments such as BICEP2 and the Keck Array \cite{Ade:2014xna,Ade:2015fwj}. Inflation \cite{Starobinsky:1979ty,Guth:1980zm,Linde:1981mu,Albrecht:1982wi} is widely accepted as a probable model for the origin of our universe, one of the hallmarks of which is the production of GW (for example \cite{Starobinsky:1979ty,Rubakov:1982df}).

Sensitivity for detecting GW in the CMB have, over the years, improved constantly. Constraints on the tensor-to-scalar ratio $r$ were tightened  \cite{Komatsu:2008hk,Ade:2014xna,Ade:2015fwj,Hinshaw:2012aka,Ade:2013zuv,Ade:2015xua,Ade:2015tva} and it is expected that a sensitivity level of $r\lesssim 0.03$ be reached in the near future \cite{Ahmed:2014ixy}. Furthermore one can optimistically expect the next decade to yield measurements of $r\lesssim 0.001$ or better \cite{Amendola:2016saw}.  Constant headway is also made in the model building front, as some models become less probable, while others gain dominance.

We study a class of models that were proposed by Ben-Dayan \& Brustein  \cite{BenDayan:2009kv}. These models sport, along with the ability to conform to known observable quantities such as the primordial power spectrum (PPS) scalar index ($n_{s}$), and its running ($n_{\text{\tiny{run}}}$), the generation of appreciable amplitude of GW signal. This type of models  appear in many fundamental physics frameworks, such as effective field theory, supergravity and string theory.  A discussion regarding small field models and the possibility of GW generation \cite{Hotchkiss:2011gz,Antusch:2014cpa,Garcia-Bellido:2014wfa} soon followed. In these models, high values of $r$ in the CMB are generally associated with a scale dependence of the scalar power spectrum. We study the models proposed by Ben-Dayan \& Brustein using exact calculations. For each model, we solve the background eqautions and the Mukhanov-Sassaki (MS) equations \cite{Mukhanov:1981xt,Sasaki:1983kd,Mukhanov:1985rz} to obtain a primordial power spectrum. This process is applied to a large sample of models and allows us to study the dependence of cosmological parameters on the potential parameters with unprecedented accuracy.

Significant differences between analytical predictions of the commonly used Stewart-Lyth (SL) expressions \cite{Stewart:1993bc,Lyth:1998xn} for CMB observables and the precise results were found. These discrepencies were already found in \cite{Lesgourgues:2007gp}, however all previous discussions of such models  \cite{BenDayan:2009kv,Hotchkiss:2011gz,Antusch:2014cpa,Garcia-Bellido:2014wfa} nevertheless heavily rely on the SL expression, thus the importance of this discrepancy is enhanced.  These differences arise from several factors, chief among them is breaking of slow-roll hierarchy. When the hierarchy is broken the time derivatives of the first and second slow-roll parameters ($\epsilon_H,\delta_H$) cannot be neglected. Hence, rather than general arguments, these models require precise calculations in order to study their validity. This also means that, in some cases, it is not possible to use Hankel functions as an approximate solution of the MS equation. This was discussed in some length in \cite{Wang:1997cw}. In other cases the Hankel functions can still be used, but either require adjustments, or some additional requirements must be met as in \cite{Schwarz:2001vv}.

\section{The primordial power spectrum and the cosmological parameters
\label{sec:PPS observables}}
The primordial power spectrum (PPS) is traditionally characterized by its spectral index $n_s$ and the index running $n_{\text{\tiny{run}}}$ (sometimes also denoted as $\alpha$ ), which are given by the first and second logarithmic derivatives of the logarithm of the PPS:
\begin{align} n_{s}=1+\left.\frac{\partial \log\left(P_{s}\right)}{\partial \log\left(k\right)}\right|_{aH=k},
\end{align}
\begin{align} n_{\text{\tiny{run}}}=\left.\frac{\partial^{2} \log\left(P_{s}\right)}{\partial \log\left(k\right)^{2}}\right|_{aH=k}=\left.\frac{\partial n_{s}}{\partial \log\left(k\right)}\right|_{aH=k},
\end{align}
where $aH=k$ denotes the CMB scale.

\subsection{A brief review}
The process of relating slow-roll parameters to the power spectrum is documented extensively in \cite{Stewart:1993bc} and described in broad strokes in \cite{Lyth:1998xn}.

Following is a brief review of the process.

In principle, the process of deriving the PPS given an inflationary potential is straightforward. The background evolution equations
\begin{align}
\left\{
\begin{array}{ccc}
\dot{H}&=&-\frac{\dot{\phi}^2}{2}\\
&&\\
\ddot{\phi}&=&-3H\dot{\phi}-\frac{dV}{d\phi}
\end{array}
\right.
\end{align}
are solved to construct the pump field:
\begin{align}
	Z=\frac{a\dot{\phi}}{H},
\label{Z}
\end{align}
where a dot denotes a derivative with respect to cosmic time.
The MS equations \cite{Mukhanov:1981xt,Sasaki:1983kd,Mukhanov:1985rz} are:
\begin{align}
\frac{\partial^{2} U_{k}}{\partial \tau^{2}}+U_{k}\cdot\omega_{k}^{2}(\tau)=0,\label{MSfirst}
\end{align}
in conformal time $\tau$ and in Fourier space with wave vector $k$, where $\omega(\tau)$ is given by:
\begin{align}
	\omega_{k}^{2}(\tau)\equiv\left(k^{2}-\frac{Z''}{Z}\right)\;.
\label{omega}
\end{align}
Here a prime denotes a derivative with respect to conformal time. The eigenfunctions $U_{k}(\tau)$ of these equations are recovered. Evaluating these at a time $\tau$ later than the latest freeze-out time yields the PPS generated by the inflationary potential $V$.
\begin{figure}[!ht]
\includegraphics[width=1\textwidth]{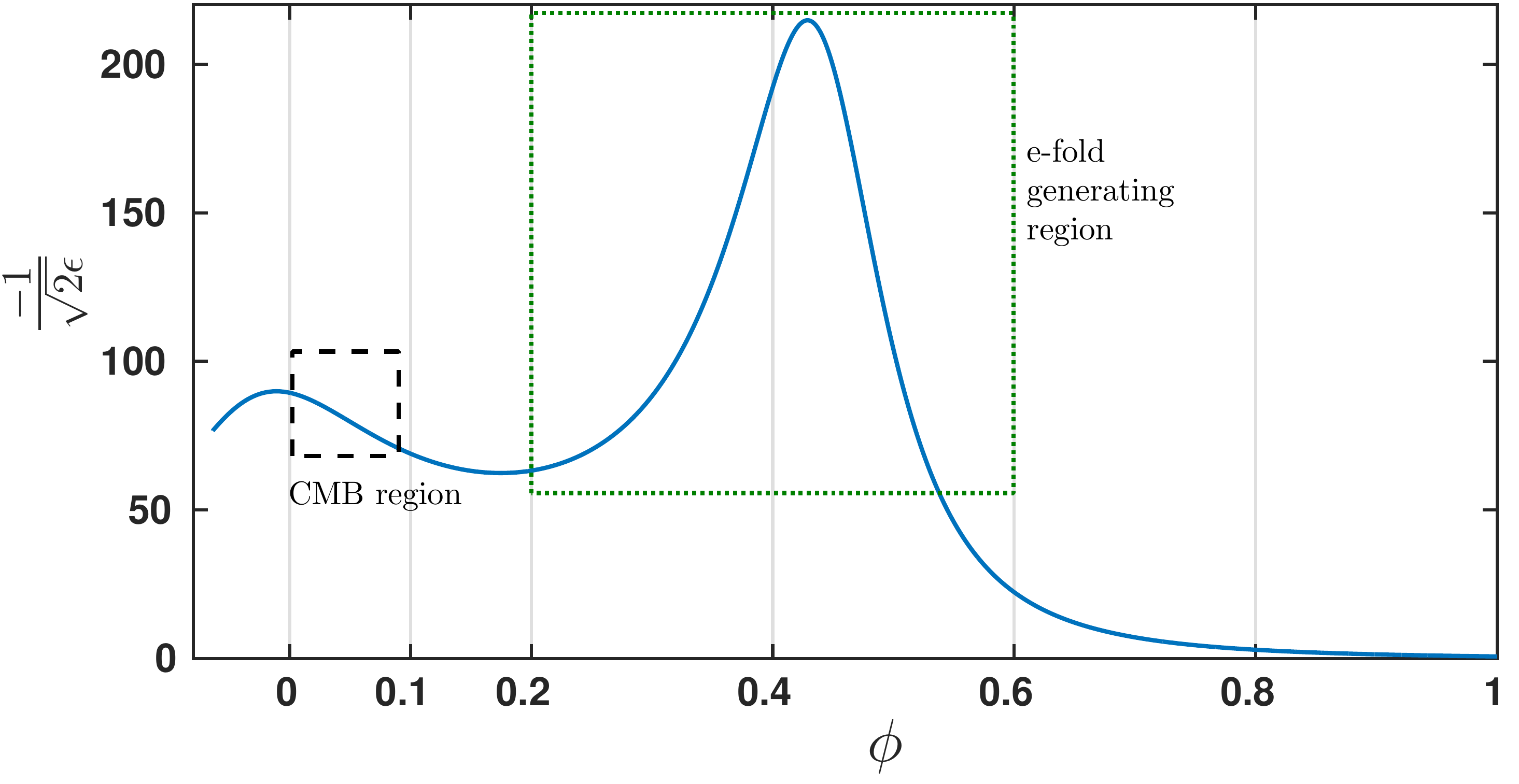}
\caption{A graph depicting $-1/\sqrt{2\epsilon}$ as a function of the inflaton $\phi$ for a model for which $r_0=0.001$. The CMB interval is covered by $\sim 8$ e-folds generated while the field changes by about $\Delta\phi\sim 0.1$. Most of the e-folds are generated when $\phi$ reaches $\sim 0.4$.}
\label{fig:1OverSqrt(2eps)}
\end{figure}
In  \cite{Stewart:1993bc}, Stewart \& Lyth derive an analytic expression for the spectral index of a wide array of inflationary scenarios. They first assume a slow-roll inflation, sufficiently slow, so that both slow roll parameters,
\begin{align}
	\nonumber \epsilon_{H}\equiv -\frac{\dot{H}}{H^{2}},\\
	\delta_{H}\equiv\frac{\ddot{\phi}}{H\dot{\phi}},
\end{align}
can be approximated by constants.
It is useful to rewrite the quantity $\frac{Z''}{Z}$ as:
\begin{align}
\frac{Z''}{Z}=2a^{2}H^{2}\left[1+\frac{3\delta_H}{2}+ \epsilon_H+\frac{\delta_H^{2}}{2}+\frac{\epsilon_H\delta_H}{2}+\frac{1}{2H} \left(\dot{\epsilon_H}+\dot{\delta_H}\right)\right].
\label{Z''/Z}
\end{align}
For strictly constant $\epsilon_H$, $\delta_H$,
\begin{align}
	\frac{Z''}{Z}=\frac{\widetilde{C}}{\tau^2},
\end{align}
with $\widetilde{C}$ a constant. In this case, the background solution corresponds to power law inflation. The resulting MS equations becomes the Bessel equations which can be solved analytically. When the Bunch-Davies boundary conditions are imposed, the resulting solution is given by a Hankel function of the  first kind:
\begin{align}
U_{k}(\tau)=\sqrt{\frac{\pi}{4}} e^{i(\nu+\frac{1}{2})\frac{\pi}{2}}\sqrt{-\tau}\mathcal{H}^{(1)}_{\nu}(-k \tau), \label{HankelSolution}
\end{align}
with the index $\nu$ given by:
\begin{equation}
\nu=\frac{3+2\delta_H+\epsilon_H}{2(1-\epsilon_H)}.
\label{nuindex}
\end{equation}
The resulting power spectrum is given by:
\begin{align}
(P_{R})^{\frac{1}{2}}=2^{\nu-\frac{3}{2}} \left(1-\epsilon_H\right)^{\frac{1}{2}-\nu} \frac{\Gamma(\nu)}{\Gamma\left(\frac{3}{2}\right)}\frac{H^{2}}{2\pi |\dot{\phi}|}.
\end{align}
Upon derivation of $P_R$  with respect to $\log(k)$ and evaluating the derivative  at $k= aH$ the scalar index is obtained,
\begin{align}
	\nonumber \frac{n_{s}-1}{2}=& \\ \nonumber &\frac{\partial \nu}{\partial \log(k)}\left[b-\log(1-\epsilon_H)\right] \\ &-\frac{1-2\nu}{1-\epsilon_H}\epsilon_H(\epsilon_H+\delta_H) -2\epsilon_H -\delta_H.
\end{align}
Inserting $\nu$ from Eq.~(\ref{nuindex}) one gets:
\begin{align}
\nonumber n_{s}-1&= \\ \nonumber &+2\times\left[\left(4\epsilon_H^{2}+ 5\epsilon_H\delta_H-\delta_H^{2}+ \frac{\delta_H\dddot{\phi}}{H\ddot{\phi}}\right)\left(b-\log(1-\epsilon_H)\right)\right.\\ &\left.-2\left(\frac{\epsilon_H}{1-\epsilon_H}\right) (1+2\epsilon_H+\delta_H)(\epsilon_H+\delta_H)-2\epsilon_H-\delta_H\right],
\label{nsH}
\end{align}
with $b=2-\log(2)-\gamma$, $\gamma$ being the Euler number. The resulting scalar index running is given (for instance in \cite{Lyth:1998xn}) by:
\begin{align}
n_{\text{\tiny{run}}}=-8\epsilon_H^{2}-10\epsilon_H\delta_H +2\delta_H^{2}-2\frac{\delta_H\dddot{\phi}}{H\ddot{\phi}}.
\label{nrunH}
\end{align}
We note that
\begin{align*}
	\frac{\dot{\delta_H}}{H}=\delta_H\left(\epsilon_H -\delta_H +\frac{\dddot{\phi}}{H\ddot{\phi}}\right)\simeq \frac{\delta_H\dddot{\phi}}{H\ddot{\phi}}-\delta_H^2,
\end{align*}
which in the slow-roll paradigm is usually taken to be small, appears in both Eq.~(\ref{nsH}) and Eq.~(\ref{nrunH}). It might be tempting then, to drop these terms. However, this term was shown in \cite{Stewart:1993bc}, and later in \cite{Lyth:1998xn} to be required for a better than $\sim 1\%$ accurate prediction of the CMB observables.
\begin{figure*}[!h]
\includegraphics[width=1\textwidth]{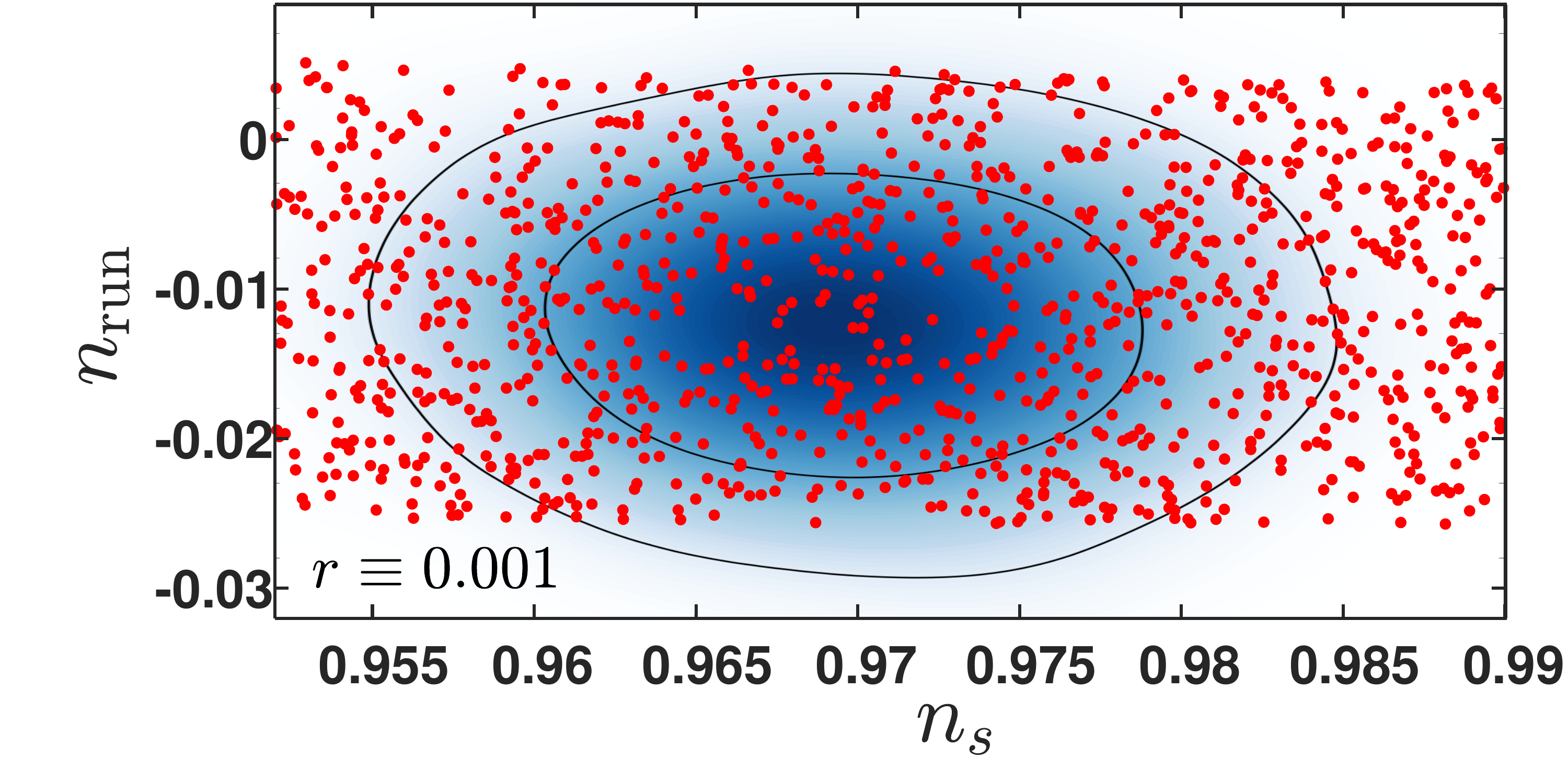}
\caption{Shown are the results of evaluating $n_s$ and $n_{\text{\tiny{run}}}$ for about 1100 models for which $r_0=0.001$.  The contour curves are the $68\%$ and $95\%$ confidence estimators, obtained from a CosmoMC  $\Lambda$CDM + index running model run \cite{Lewis:2002ah} using the Planck \& Bicep joint data analysis \cite{Ade:2015tva}.  The pivot scale used in the analysis is $k_{pivot}=0.05\; h\; Mpc^{-1}$, which is the same scale as in \cite{Ade:2015tva}.  \label{figure:FirstResult}}
\end{figure*}
The authors of \cite{Stewart:1993bc} then proceed to connect slow-roll parameters to the potential and its derivatives by a process of Taylor expanding with respect to cosmic time, and re-substituting the Friedman equations to $2^{nd}$ order. Thus they are able to obtain an analytical expression that connects the PPS observables directly to the potential and its derivatives to a high degree of accuracy. Following the same procedure for the running of the scalar index, yields (again, to $2^{nd}$ order):
\begin{align}
	n_{s}\simeq&1-6\varepsilon_{V,0}+2\eta_{V,0}
\label{nsV} \\
	&\nonumber +2\Bigg[\frac{\eta_{V,0}^{2}}{3}- \left(\frac{5}{3}-12b\right)\varepsilon_{V,0}^{2} \\ \nonumber &- \left(8b+1\right)\varepsilon_{V,0}\eta_{V,0} +\left(b+\frac{1}{3}\right)\xi_{V,0}^{2}\Bigg] ,\\ \nonumber \\
	n_{\text{\tiny{run}}}\simeq & 16\varepsilon_{V,0}\eta_{V,0} -24\varepsilon_{V,0}^{2}-2\xi_{V,0}^{2},
\label{nrunV}
\end{align}
where the subscript ${0}$ denotes evaluating the quantity at the CMB point and the subscript $V$ denotes  that these are potential derivatives:
\begin{align}
	\varepsilon_{V}=\frac{1}{2}\left(\frac{V'}{V}\right)^{2},\\
	\eta_{V}=\frac{V''}{V},\\
	\xi_{V}^{2}=\frac{V'V'''}{V^2}.
\end{align}
However, when $\epsilon_H$ or $\delta_H$ are not strictly constants, the analytic solution to the MS equation is not generally known. We show that the above analytic expressions are not accurate enough for certain models where $\epsilon_H$ and $\delta_H$ are time-dependent. Therefore, one has to use the precise calculations which takes into account the deviations of the MS equation solutions from the Hankel functions.

Another observable, used to parametrize the amplitude of GW at the onset of inflation is the scalar-to-tensor ratio $r$,  $r=16\epsilon_H $. In fact, should we ever detect a GW signal, we would be able to directly probe the energy scale of inflation \cite{Lyth:1996im}.
\begin{figure*}[!h]
\includegraphics[width=1\textwidth]{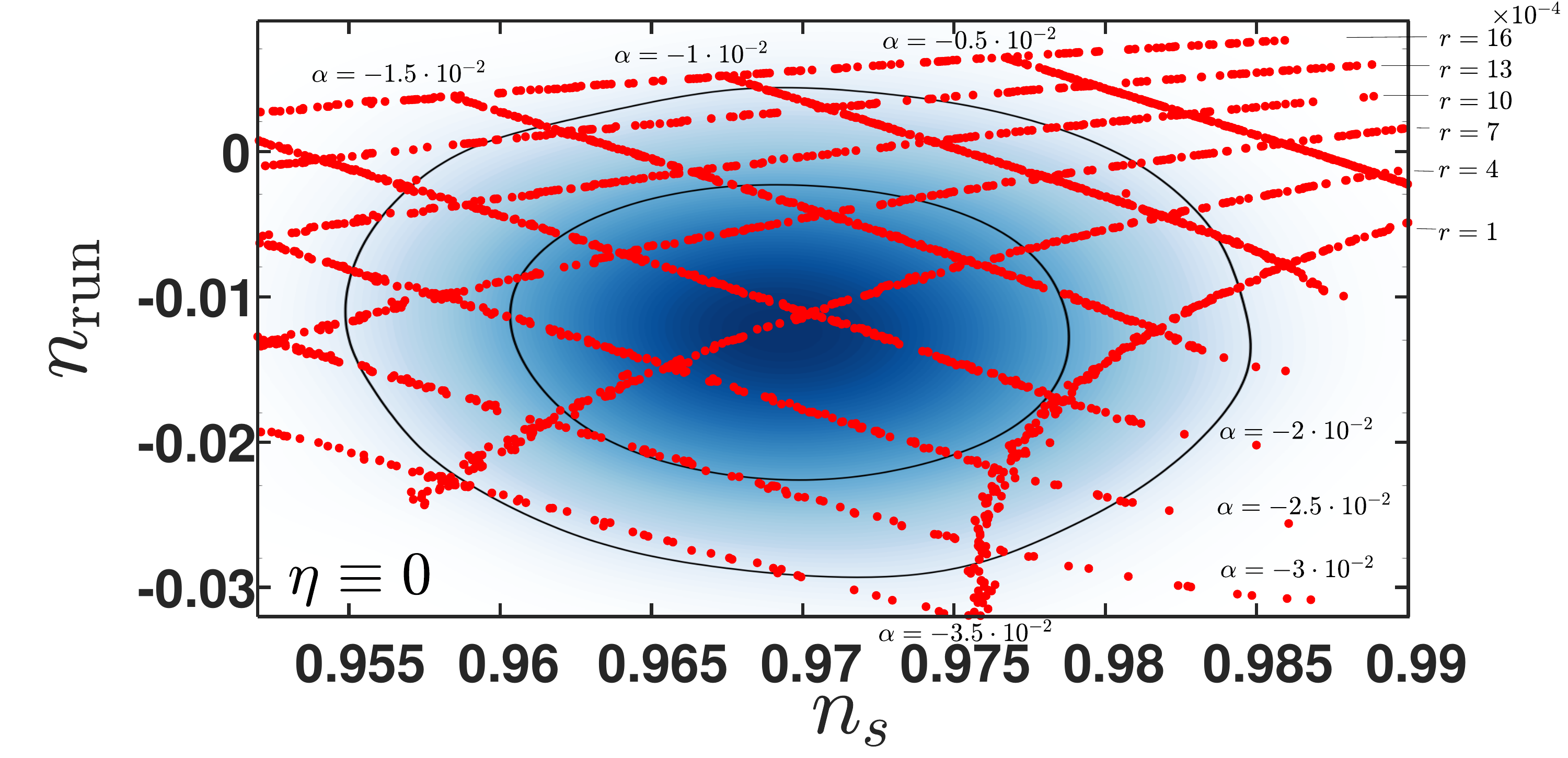}
\caption{Covering the $n_s-n_{\text{\tiny{run}}}$ plane with constant $r$  and constant $\alpha$ characteristics, for $\eta_0 =0$}\label{figure:SecondResult}
\end{figure*}
\section{Inflationary models}
\label{sec:smallField}

Small field models of inflation in which inflation occurs near a flat feature, a maximum, or a saddle point are studied (see \cite{Boubekeur:2005zm} for a review). This class of models is interesting because they appear in many fundamental physics frameworks, effective field theory, supergravity \cite{Yamaguchi:2011kg} and string theory \cite{Baumann:2014nda} in successive order of complexity. Our focus on such models is also motivated by the expected properties of the moduli potentials in string theory. More generally speaking these type of models can be viewed as a Taylor expansion approach to other models \cite{Dodelson:1997hr}. A different more observable-oriented classification of models can be found in \cite{Schwarz:2004tz}, in which analysis our models fall into the toward-exit class.

In general, inflation will occur in a multi-dimensional space, however, the results for multifield inflation cannot usually be obtained in a simple way. In many known cases it is possible to identify a-posteriori a single degree of freedom along which inflation takes place. To gain some insight about the expected typical results effective single field potentials can be used.

Generic small field  models predict a red spectrum of scalar perturbations,  negligible spectral index running and non-gaussianity. They also predict a characteristic suppression of tensor perturbations \cite{BenDayan:2008dv}. Hence, they were not viewed as candidate models for high-$r$ inflation. Large field models of inflation are thus the standard candidates for high-$r$ inflation.

In \cite{BenDayan:2009kv}, a new class of more complicated single small field models of inflation was considered (see also \cite{Hotchkiss:2011gz}) that can predict, contrary to popular wisdom \cite{Lyth:1996im,Martin:2013tda}, an observable GW signal in the CMB (see also \cite{Cicoli:2008gp}.) The notion that observable signal GW precludes small field models partly stems from \cite{Martin:2013tda} and similar analyses that study monomial potential models as small field models. The spectral index, its running, the tensor to scalar ratio and the number of e-folds were claimed to cover all the parameter space currently allowed by cosmological observations. The main feature of these models is that the high value of $r$ is accompanied by a relatively strong scale dependence of the resulting power spectrum. Another unique feature of models in this class is their ability to predict, again contrary to popular wisdom \cite{Easther:2006tv}, a negative spectral index running.  The single observable consequence that seems common to all single field models is the negligible amount of non-gaussianity.
In \cite{Lesgourgues:2007gp} the inflationary potential was Taylor-expanded up to order $4$. The approach applied in \cite{Lesgourgues:2007gp} is similar, however it seems only potentials that are monotonic in the entire CMB window were considered.

The current work yields corrected predictions of this class of models by a systematic high-precision analysis, thus providing a viable alternative to the large field-high $r$ option. The analysis of \cite{BenDayan:2009kv} is extended, in preparation for a future detailed comparison of the models to data. This is done in order to simplify the parametrization of the potential and facilitate a comprehensive numerical study.
\subsection{Inflaton potentials}
The following class of polynomial inflationary potentials proposed in \cite{BenDayan:2009kv} is:
\begin{align}
V(\phi)=V_{0}\left(1+\sum_{p=1}^{5}a_{p}\phi^{p}\right)\;. \label{Full_form}
\end{align}

The virtue of these models from a phenomenological point-of-view is the ability to separate the CMB region from the region of large e-fold production. Hence, these potentials can produce  a very different spectrum early on, than in the later stages of inflation. Figure \ref{fig:1OverSqrt(2eps)} illustrates this point, with separate CMB region and e-fold generation region.
In the context of both classification systems mentioned, current observational data weakly support these \cite{Martin:2014lra,Vennin:2015eaa}. However the small field model studied in \cite{Martin:2014lra} are monomial potential models of the form $V\propto 1-a_p\phi^p$, which are different from many of our models.

In many models $\varepsilon_V\sim 1/N^2$, $\eta_V\sim 1/N^2$, and the time derivative $\frac{d}{Hdt}$ can approximately be replaced with a factor of $\frac{1}{N^2}$ \cite{Kosowsky:1995aa}. In the above models this standard hierarchal dependence is broken, they have a more complicated dependence while obeying the slow-roll conditions $\epsilon_H$, $\delta_H \ll 1$.
In \cite{BenDayan:2009kv} it was shown that these models can be written as:
\begin{align}
V(\phi)=V_{0}\left(1-\sqrt{\frac{r_{0}}{8}}\phi +\frac{\eta_{0}}{2}\phi^{2} +\frac{\alpha_{0}}{3\sqrt{2 r_{0}}}\phi^{3} +a_{4}\phi^{4}+a_{5}\phi^{5}\right).\label{BB-potential}
\end{align}
Here  $r_0$, $\eta_{0},\alpha_{0}$ are defined as $r= 8 \left(\frac{V'}{V}\right)^2$,  $\eta=\frac{V''}{V}$, $\alpha=-2\xi^{2}$, respectively. The subscript $0$ means that these are the values at the CMB point.

Specifically for a potential of the form $V\propto 1+\sum_{p=1}^5a_{p}\phi^p$, the SL analytic expression for the scalar index and its running (Eq.~(\ref{nsV},\ref{nrunV})) is given by
\begin{align}
    n_{s}\simeq&1-3a_{1}^2+4a_{2} \\
    &\nonumber +2\Bigg[\frac{4a_{2}^2}{3}-\left(\frac{5}{3}-12b\right)\frac{a_{1}^4}{4}\\ \nonumber &-\left(8b+1\right)a_{1}^2 a_{2} +\left(6b+2\right)a_{1}a_{3}\Bigg] ,\\ \nonumber \\
    n_{\text{\tiny{run}}}\simeq & 16a_{1}^2 a_{2}-6a_{1}^4-2a_{1}a_{3}.
\end{align}

\subsection{Reduced parameter space}
\begin{figure}[!h]
\includegraphics[width=1\textwidth]{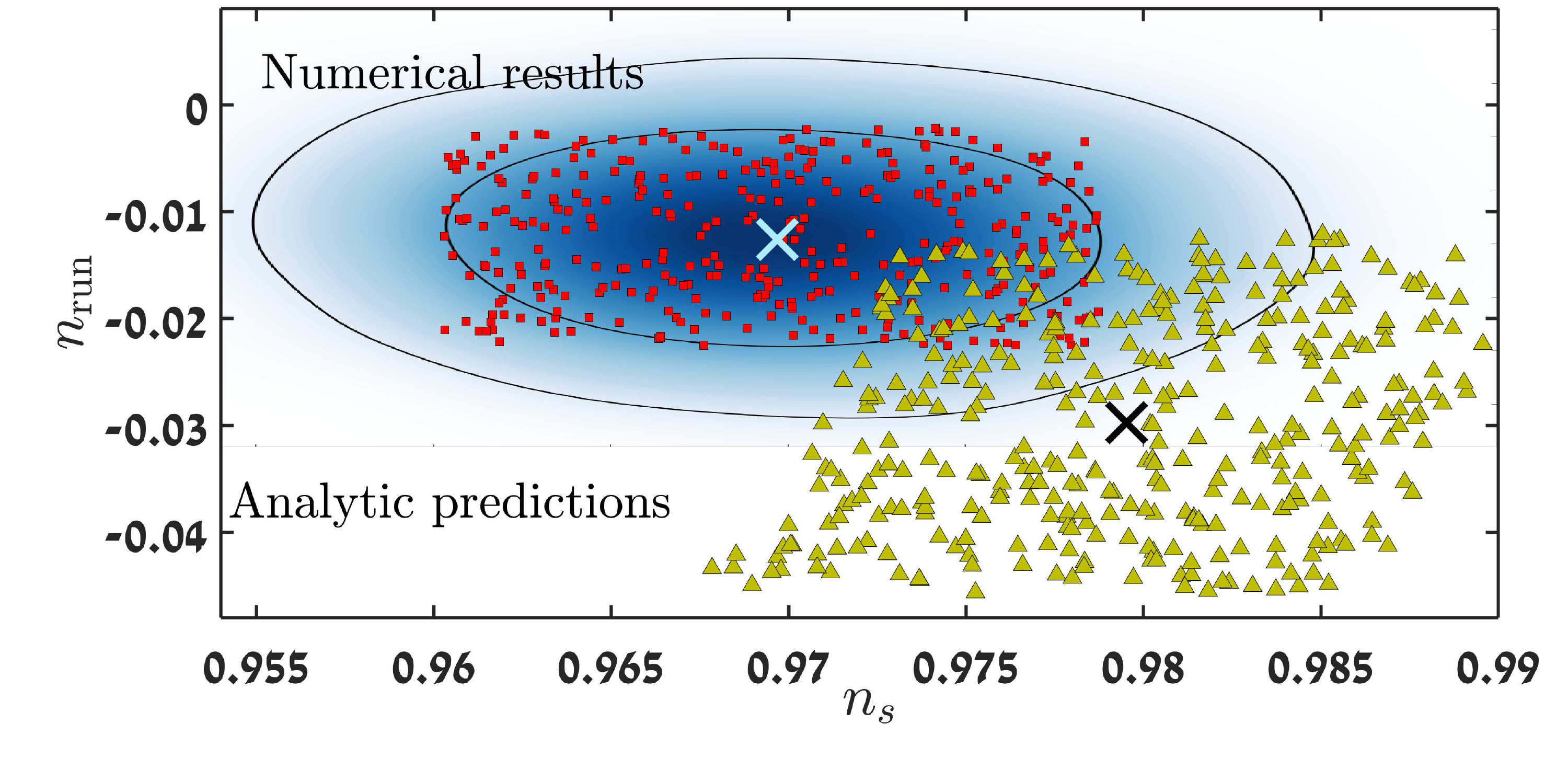}
\caption{Shown are the results of a precise calculation of the cosmological parameters of $\sim$ 200 models (red squares), as well as the corresponding analytic predictions (yellow triangles) calculated according to \eqref{nsV} and \eqref{nrunV}. The cyan and black x's  mark the mean value of the precise and analytic results (respectively).}\label{fig:Errors}
\end{figure}
The potential in \eqref{BB-potential} is a small field candidate, which after some scaling and normalization, depends on four free parameters. One parameter is used for setting  $r_0$ at the CMB point, and thus the predicted amplitude of the GW signal produced, while the other two parameters are used to parametrize the $n_s-n_{\text{\tiny{run}}}$-plane. The fourth parameter determines the number of e-folds from the CMB point to the end of inflation. $\phi_{\text{\tiny{end}}}$ is set to $1$ to simplify the analysis. It follows that
 \begin{align}
  \frac{1}{2}\left(\frac{V'}{V}\right)^2{|_{\phi=1}}=1. \label{CONDITION}
\end{align}
Suppose we want inflation to end at $\phi=\alpha$, we can rescale $\phi$:
	\begin{align}
		\phi\rightarrow \tilde{\phi}=\frac{\phi}{\alpha}.
	\end{align}
	In this formulation,
	\begin{align}
		V=V_0\left(1+\sum_{p}a_p \alpha^{p} \tilde{\phi}^{p}\right)=V_0\left(1+\sum_{p} \tilde{a_p}\tilde{\phi}^p\right),
	\end{align}
	where $\tilde{a_p}=a_p \alpha^{p}$. Since this is the exact same potential, it follows the exact same CMB observables are yielded. Thus, applying condition \eqref{CONDITION} can be viewed as a scaling scheme for the different terms in the potential which does not limit the generality of our results. \\
	
~ Substituting the expression for the potential and its derivative at $\phi=1$ we get:
\begin{align}
	-\sqrt{2}=\frac{\sum_{p=1}^{5}p\cdot a_p}{1+\sum_{p=1}^{5}a_p}.
\end{align}
 $a_4$ is now given in terms of the other coefficients:
 \begin{align}	
a_4=\frac{-1}{4+\sqrt{2}}\left(\sqrt{2} +\sum_{p\in(1,2,3,5)}\left(p+\sqrt{2}\right)a_p\right)
 \label{sola4}
 \end{align}

Using the standard definition for the number of e-folds  $N=\int_{\phi_{\text{\tiny{CMB}}}}^{\phi_{\text{\tiny{end}}}}Hdt\simeq -\int_{\phi_{\text{\tiny{CMB}}}}^{\phi_{\text{\tiny{end}}}}\frac{V}{V'}d\phi$, and the approximation $V(\phi)=1+\sum_{p=1}^{5}a_{p}\phi^{p}\simeq 1$ yields a rough estimate for  $a_5$ as a function of $N$,
\begin{align}
N\simeq-\int_{0}^{1}\frac{V(a_{1},a_{2},a_{3},a_{5})}{V'(a_{1},a_{2},a_{3},a_{5})}d\phi \simeq -\int_{0}^{1}\frac{d\phi}{V'(a_{1},a_{2},a_{3},a_{5})}.
\end{align}
This estimate is then used as a starting point to refine $a_5$ by solving the background equations iteratively thereby obtaining the accurate coefficient $a_5$ that yields the correct $N$. Thus a 4-dimensional parameter space $r_0$, $a_2$, $a_3$, $N$ is defined. The parameters  $a_{2},a_{3}$ are constrained by the requirement $|a_{2}|,|a_{3}|\ll 1$, $a_{1}$ is constrained by the observable value of $r$ and $a_{5}$ is determined by the other parameters and by the number of e-folds (taken to be in between $50\sim 60$).  The PPS considered is in the range of the first $\log(2500)\sim 8$ e-folds of inflation.

\begin{figure}[!ht]
 \includegraphics[width=1\textwidth]{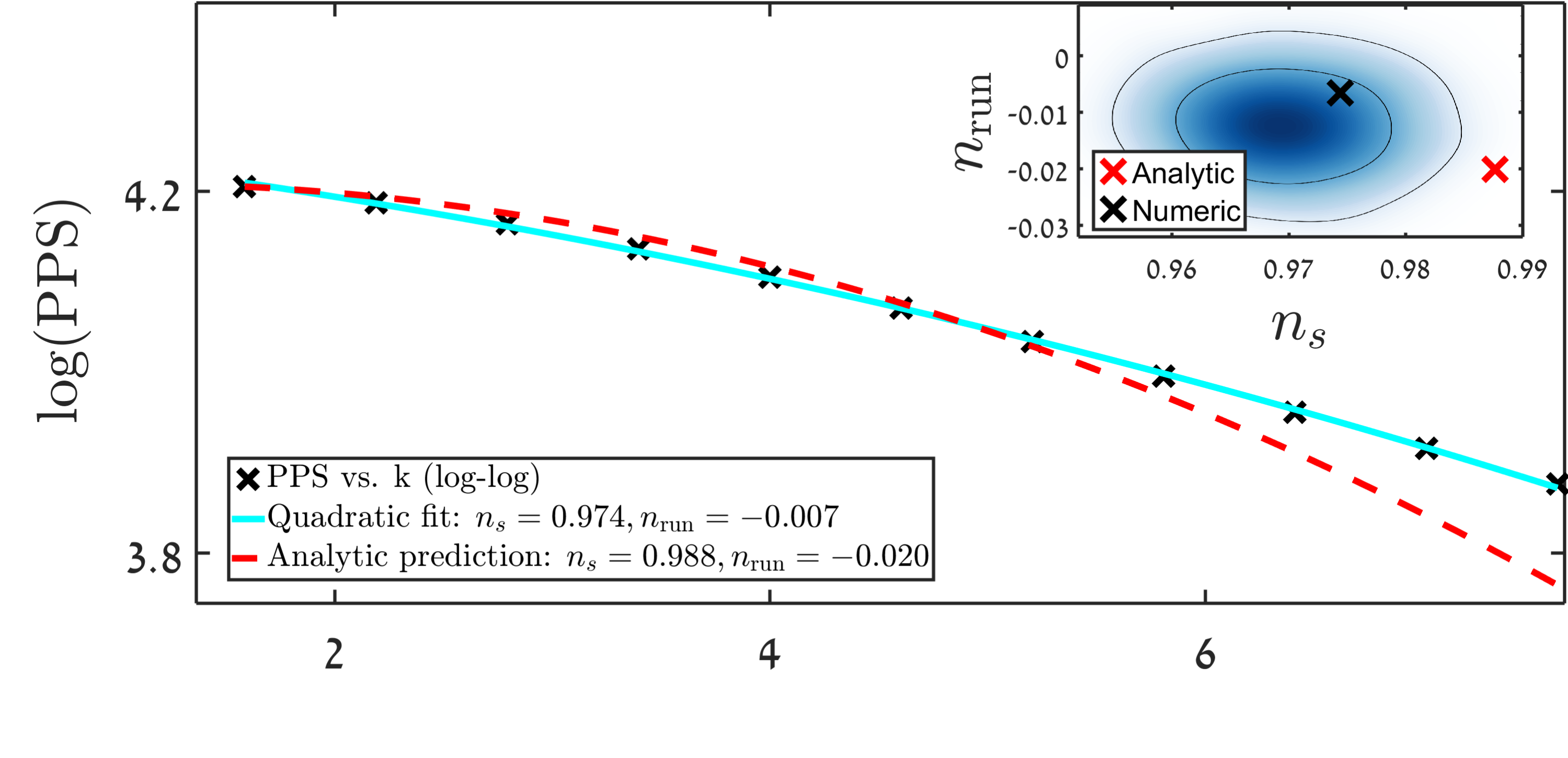}\\
  \includegraphics[width=1\textwidth]{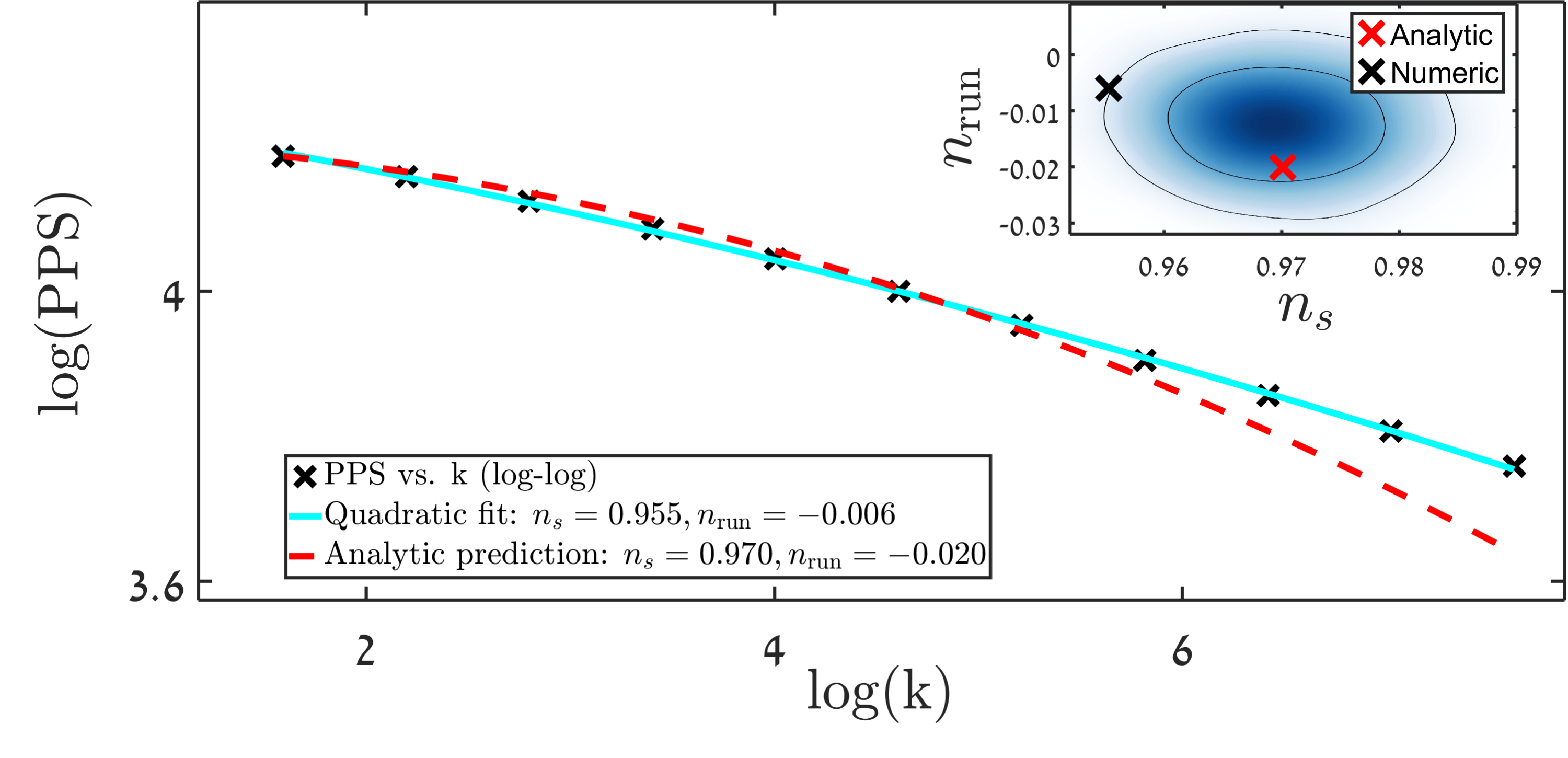}
  \caption{ Comparison of the precise results and analytic predictions made with \eqref{nsV}. Each panel shows the precisely calculated results, fitted by a quadratic polynomial to extract $n_{s}$ and $n_{\text{\tiny{run}}}$.  The curve predicted by \eqref{nsV} is plotted as a reference.  In the upper panel we show a potential that would be excluded based on the analytic result, whereas the precise results is well within the $68\%$ probability curve. In the lower panel the exact opposite is the case, with an analytically accepted result, but an excluded precise one.}\label{fig:exemplars}
\end{figure}
\section{Precise evaluation of the cosmological parameters}

Using the analytic results in Eqs.~(\ref{nsV}) and (\ref{nrunV}) it can be concluded that the above class of models  can cover the part of the $n_{s}-n_{\text{\tiny{run}}}$ plane of interest \cite{BenDayan:2009kv}. However, several approximations are made along the way. Significant deviations from analytic prediction are found, of the order of a percent or so in estimating $n_s$ and 50\% or more in estimating $n_{\text{\tiny{run}}}$. The unavoidable conclusion is that rather than a general argument, a precise calculation is necessary to extract the cosmological parameters these models yield.

\subsection{From potentials to cosmological parameters}
The process of calculating the cosmological parameters for a given potential is the following.
A potential candidate is built by setting a parameter (for instance $r_0$), and randomly drawing the other parameters (in this example $a_2$ and $a_3$) from a uniform distribution function. The limits of this distribution function are set by hand and require a  process of trial and error (guided by theoretical insights such as overall behaviour of precisely calculated $n_s$ and $n_{\text{\tiny{run}}}$). After the 3 first parameters are fixed, Eq.~(\ref{sola4}) is used to relate $a_4$ to $a_5$, and the value of $a_5$ is calculated for a the desired value of $N$. $a_5$ is found as explained above, with no approximations. The choice of which parameters to fix and which to randomly draw relies on the observables studied.

For each potential the Friedmann equations and the inflaton scalar field equation are solved. The initial conditions are set such that integration starts 3.5 efolds before the CMB point with $\dot{\phi}=0$. In that fashion we ensure that we are well within the slow roll regime, and on the attractor solution when the CMB point is reached. The solution is used to construct $Z$ and $\omega_{k}$ as described in \eqref{Z} and \eqref{omega}. The eigenfunctions for the MS equations \eqref{MSfirst} are found and used to calculate the power spectrum. Finally we provide a fit for the power spectrum, from which we extract $n_{s}$ and $n_{\text{\tiny{run}}}$.
\section{Cosmological parameters of small field models}
\begin{table}[!h]
\begin{center}
\begin{tabular}{||c|c|c|c|c|c||c||}
\hline
$a_2$&$a_3$&precise&analytic&precise&analytic&Fit error\\
&&$n_s$&$n_s$&$n_{\text{\tiny{run}}}$&$n_{\text{\tiny{run}}}$&  \\
&&&&&&($\times 10^{-4}$)\\
\hline
$0.0005$&$-0.3041$&$0.9777$&$0.9856$&$-0.0196$&$-0.0409$&$	1.8$\\
\hline
$-0.0013$&$-0.2795$&$0.9713$&$0.9796$&$-0.0175$&$-0.0373$&$	1.5$\\
\hline
$-0.0001$&$-0.2188$&$0.9780$&$0.9877$&$-0.0125$&$-0.0293$&$	1.1$\\
\hline
$-0.0042$&$-0.1538$&$0.9627$&$0.9748$&$-0.0067$&$-0.0203$&$	0.8$\\
\hline
$-0.0032$&$	-0.2923$&$0.9631$&$0.9711$&$-0.0185$&	$-0.0387$&$	1.9$\\
\hline
$-0.0002$&$-0.2709$&$0.9760$&$0.9843$&$-0.0168$&	$-0.0363	$&$	1.6$\\
\hline
$-0.0026$&$-0.1342$&$0.9710$&$0.9820$&$-0.0055$&	$-0.0178	$&$	0.6$\\
\hline
$-0.0031$&$-0.1517$&$0.9670$&$0.9793$&$-0.0066$&	$-0.0201	$&$	0.8$\\
\hline
$-0.0011$&$	-0.1563$&$0.9757$&$0.9868$&$-0.0072$&	$-0.0209	$&$	0.7$\\
\hline
$-0.0024$&$	-0.2808$&$0.9662$&$0.9752$&$-0.0174$&	$-0.0373	$&$	1.9$\\
\hline
\end{tabular}
\caption {Shown is a table of 10 potentials constructed such that $r_0=0.001$, and $N=60$. The parameters $a_2$ and $a_3$ are constructed by randomly drawing from a uniform distribution as explained in Section 4. The discrepancy in $n_s$ is around $0.8\%\sim 1.25\%$, while the $n_{\text{\tiny{run}}}$ discrepancy is much more pronounced.  } \label{table:PotTable}
\end{center}
\end{table}
In this section we present the results of evaluating cosmological parameters for many small field models. In Fig.~\ref{figure:FirstResult} we show an example for which we calculate $n_s$ and $n_{\text{\tiny{run}}}$ for about 1100 models with a fixed scalar to tensor ratio $r_0=0.001$. The results are shown on a $n_{s}-n_{\text{\tiny{run}}}$ joint probability graph with the $68\%,95\%$ contours that are the probability estimators as yielded by a CosmoMC \cite{Lewis:2002ah} $\Lambda$CDM +index running model run, with the most recent Bicep \& Planck data (including WMAP 9-year mission)  \cite{Ade:2015tva}.

The reason for choosing the value of $r_0=0.001$ (and not a higher value, for example, $r=0.01$) was the following. We discovered that as we increased the values of $r$, the inflaton potentials needed to be more complicated and additional parameters were required. Also, we encountered several technical difficulties which we were able to resolve for the lower values of $r$.  Solving these difficulties and constructing a reliable framework for numerical calculations of the CMB observables is an essential step towards building models with higher values of $r$, which we intend to do in a future publication.

We allow the values of $n_s$ to vary quite substantially, rather than restrict them to the narrow range that is allowed by the data. Our idea is that when $r$ and $n_{\text{\tiny{run}}}$  are free to vary, the constraints on $n_s$ are relaxed in a significant way. The reason is that there is some degeneracy among the parameters. This is validated in the preliminary analysis that we present in this paper. In addition, despite of the fact that some models have yielded an almost flat (and some even a blue) $n_s$ and therefore are in conflict with the data, we find their analysis useful because insight regarding the departure of precisely calculated results from what the analytic SL term ( \eqref{nsV}) predicts (see below), is gained.
\begin{figure}
\includegraphics[width=0.85\textwidth]{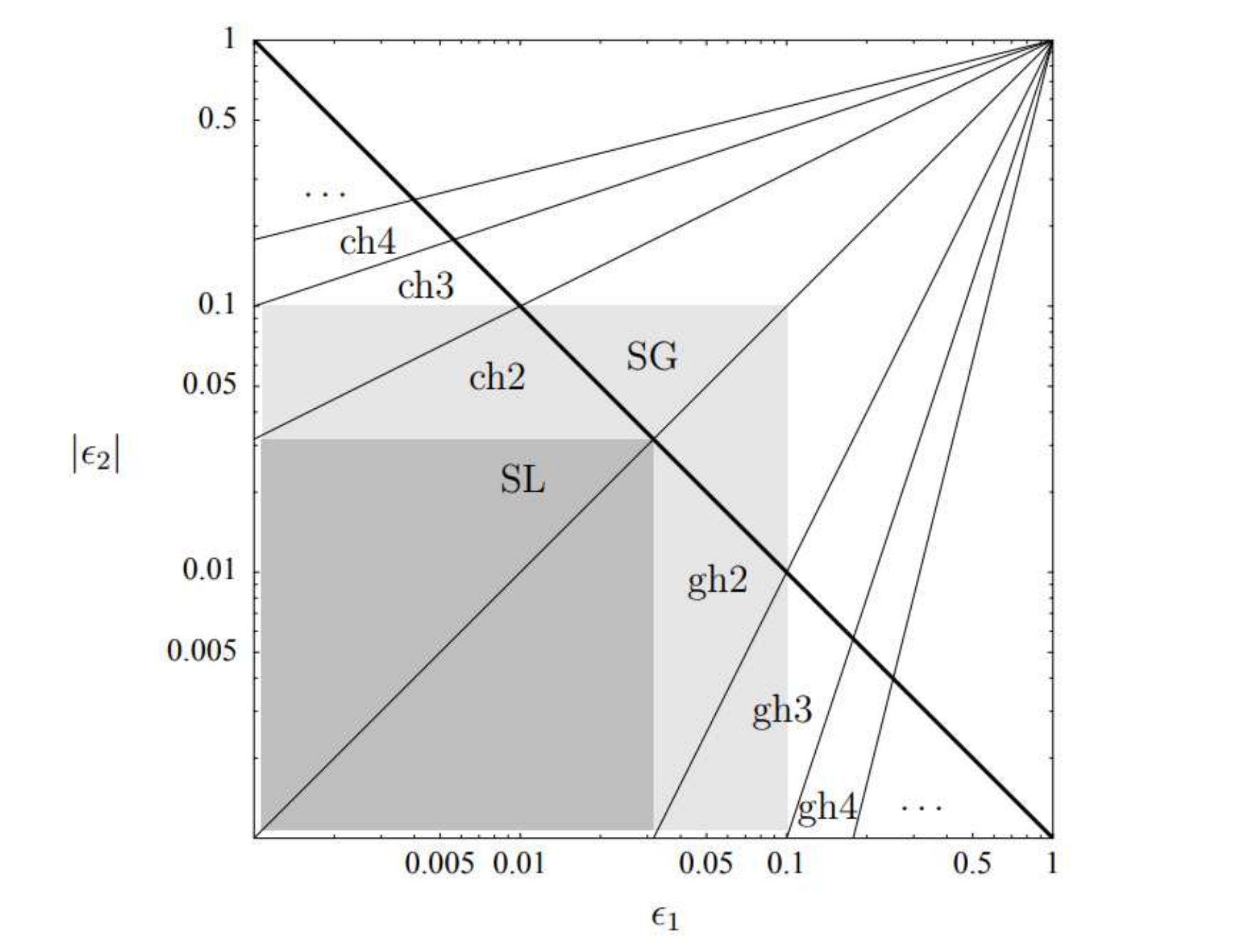}
\caption{Regions in the $\epsilon_1$-$|\epsilon_2|$ parameter space where the 
spectral amplitudes could be calculated with an accuracy better than $1\%$, according to analysis presented in \cite{Schwarz:2001vv}.
In the dark shaded region the Stewart-Lyth (SL) approximation \cite{Stewart:1993bc}, as 
well as all other approximations are supposedly sufficiently accurate. Second-order corrections,
as calculated by Stewart and Gong (SG) \cite{Gong:2001he }, extend that region to the 
light shaded region. The constant horizon approximation at order $n$ (ch$n$), 
and the growing horizon approximation at order $n$ (gh$n$), do well below
the thick line. The rays indicate where the corresponding higher order 
corrections are necessary.
The thick line itself is the condition $\epsilon_1 |\epsilon_2| < 
(A/100\%)/\Delta N$, with $\Delta N =10$ and $A = 1\%$. We study these approximations and others, and find that our models defy these analyses.
Figure and caption adapted from \cite{Schwarz:2001vv}.}\label{Different_regions}
\end{figure}
\section{Inflationary models}
\label{sec:smallField}
\subsection{Evaluating cosmological parameters for fixed $r_0$}
The $n_{s}-n_{\text{\tiny{run}}}$ plane was covered with models which yield a fixed value of $r_0=0.001$. The cosmological parameters of some 3500 potentials were calculated. Figure \ref{figure:FirstResult} shows cosmological parameters of $\sim$ 1100 models. A significant number of the models yield values of $n_{s}$ and $n_{\text{\tiny{run}}}$ within the $68\%$ and $95\%$ likelihood region. The most probable value for $\frac{V''}{V}=-0.0052 \pm 0.0034$. This is within the $68\%$ CL Planck results, with or without including high-$l$ polarization data. The third coefficient values are given by $\frac{V'''V'}{V^2}=0.0138 \pm 0.0065 $, which is in better agreement with the result without high-$l$ data. However the 2015 Planck analysis \cite{Ade:2015lrj} sets $\epsilon_4\equiv 0$ which might bias the results slightly. In the 2013 analysis \cite{Planck:2013jfk} this was not done, and our results agree with their analyses, including our values for $\frac{V^{(4)}V'}{V^2}$. Additional factors that contribute to the difference in analyses, are the approximate connection between Hubble flow functions $\epsilon_i$ and the potential derivative quantities $\epsilon_V, \eta_V,\xi^2_V$. An interesting feature of these models is the departure of precisely calculated results from what the analytic SL expression \eqref{nsV} predicts, to be discussed later. It might be possible to cover the $n_{s}-n_{\text{\tiny{run}}}$ allowed region with models with a higher scalar-to-tensor ratio. However the treatment of models which yield higher $r$ is  more complex, since by increasing $r$, one is forced to consider a larger $\Delta\phi$ range CMB region. The CMB region (see Figure \ref{fig:1OverSqrt(2eps)}) is roughly 3 times larger in $\phi$ for models with $r_0=0.01$, thus it will typically result in a running of running of the power spectrum.

\subsection{Evaluating cosmological parameters for fixed $\eta_0$}
The effects of varying $r_0$ on the resulting power spectrum were studied. In order to do this $\eta_0$ was set to $0$ for simplicity, and the $n_{s}-n_{\text{\tiny{run}}}$ plane was covered with models of varying $r_{0}$ and $\alpha_{0}$. Figure \ref{figure:SecondResult} shows the results of this study.

Notice that the effect of varying both $r_0$ and $\alpha_0$ on the changes in the value of $n_{s}$ is more pronounced than expected. Usually one expects $n_{s}-1$ to first order to be $\propto -\frac{3r_{0}}{8}$ and thus  $\Delta n_s/\Delta r_0\simeq 10^{-4}\sim 10^{-5}$. At second order, we expect $n_{s}-1$ to be $\propto \frac{\alpha_{0}}{15}$ and thus $\Delta n_s/\Delta\alpha_0\simeq 10^{-3}$, whereas in this case the change in $n_{s}$ is of the order of $10^{-2}$. A possible explanation to this phenomenon is a discrepancy between the analytic predictions made using \eqref{nsV} and the precise calculations (see below).
\begin{figure}[!ht]
\includegraphics[width=0.95\textwidth]{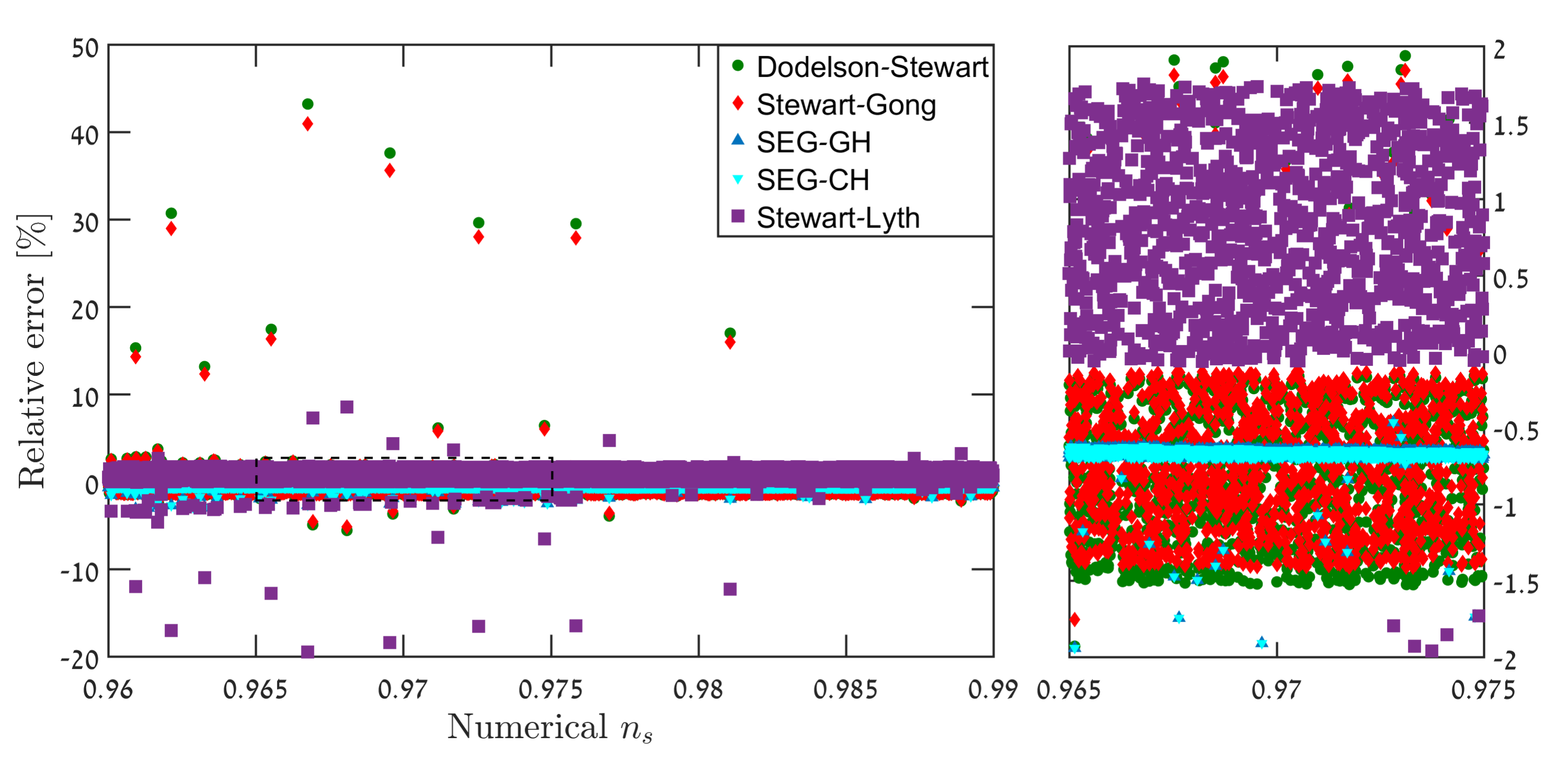}
\caption{Around 50,000 of our models numerically simulated and compared to different analytical expressions reveals a varying level of accuracy in predicting the correct scalar index. The figure shows only a partial sample of $\sim 8000$ restricted to $\epsilon_1 < 0.0275$, $|\epsilon_2| < 0.0275$ and $0.96<n_s<0.99$. Each data point is a relative error between the numerical result of a model and an analytical expression from \cite{Dodelson:2001sh} (DS,green circles), \cite{Gong:2001he} (SG,red diamonds),\cite{Schwarz:2001vv} (SEG-GH, growing horizon variant - blue triangle, and SEG-CH, constant horizon variant - inverted cyan triangle), and the usual SL \cite{Stewart:1993bc} expression (purple squares).}
\label{different_ns_SEG}
\end{figure}
\subsection{Comparison of calculated results and the Stewart-Lyth analytic predictions}
An additional study of models with a larger value of $r$ was conducted. This was done in order to confirm the ability of this class of models to produce significant GW signal, while yielding acceptable values of $n_{s}$ and $n_{\text{\tiny{run}}}$. For $r\gtrsim 10^{-3}$, a significant deviation from  the analytical expressions in Eqs.~(\ref{nsV}) and (\ref{nrunV}) was found. Potentials that by the standard analytic treatment should have yielded acceptable observables, were wide off the mark. On the other hand potentials which were supposed to be ruled out, yielded observables inside the $n_{s}-n_{\text{\tiny{run}}}$ acceptable domain. Figure~\ref{fig:exemplars} elucidates this point, with a potential (Fig.~\ref{fig:exemplars}, upper panel), for which $r_0=0.001$. The resulting $n_s$ and $n_{\text{\tiny{run}}}$ are within the $68\%$ probability allowed region, while the analytic expressions yield values outside the $95\%$ probability allowed region. The example in Fig.~\ref{fig:exemplars}, lower panel shows the opposite also occurs.

Table \ref{table:PotTable} contains as examples ten specific potentials that were chosen such that the precise results for $n_s$ and $n_{\text{\tiny{run}}}$ are within the accepted values. All of the models  produce a tensor to scalar ratio $r_0=0.001$.   The table also contains the analytic predictions made with Eqs. ~(\ref{nsV},\ref{nrunV}). As can be seen from the table, the discrepancy between the analytic predictions and the precise calculations can be quite significant for $n_{\text{\tiny{run}}}$. The spectrum is composed using 15 $k-mode$s, and the error in the rightmost column is the cumulative error defined as $error=\sqrt{\sum_k\left(fit(\log k)-sample(\log k)\right)^2}$. The mean deviation per $k-mode$ is the error divided by 15.
The differences between the analytic predictions and the results of precise calculations are quite common for this type of inflationary potentials for $r\gtrsim 0.001$, as shown in Fig.~\ref{fig:Errors}. About 3500 potentials were analysed (Fig.~\ref{fig:Errors} show only a partial sample), and $n_{s}$ and $n_{\text{\tiny{run}}}$ were extracted for each. The deviation in $n_{s}$ between analytic predictions and precise results, normalized by the sum of the two is then found. Fig.~\ref{fig:Errors} also shows a marked drift towards lower values of $n_{s}$ and higher values of $n_{\text{\tiny{run}}}$. The mean drift is approximately given by $(\Delta n_{s},\Delta n_{\text{\tiny{run}}})=(-0.01,0.02)$, with $\sim 17-18\%$ standard deviation.
\begin{figure}[!ht]
\includegraphics[width=0.95\textwidth]{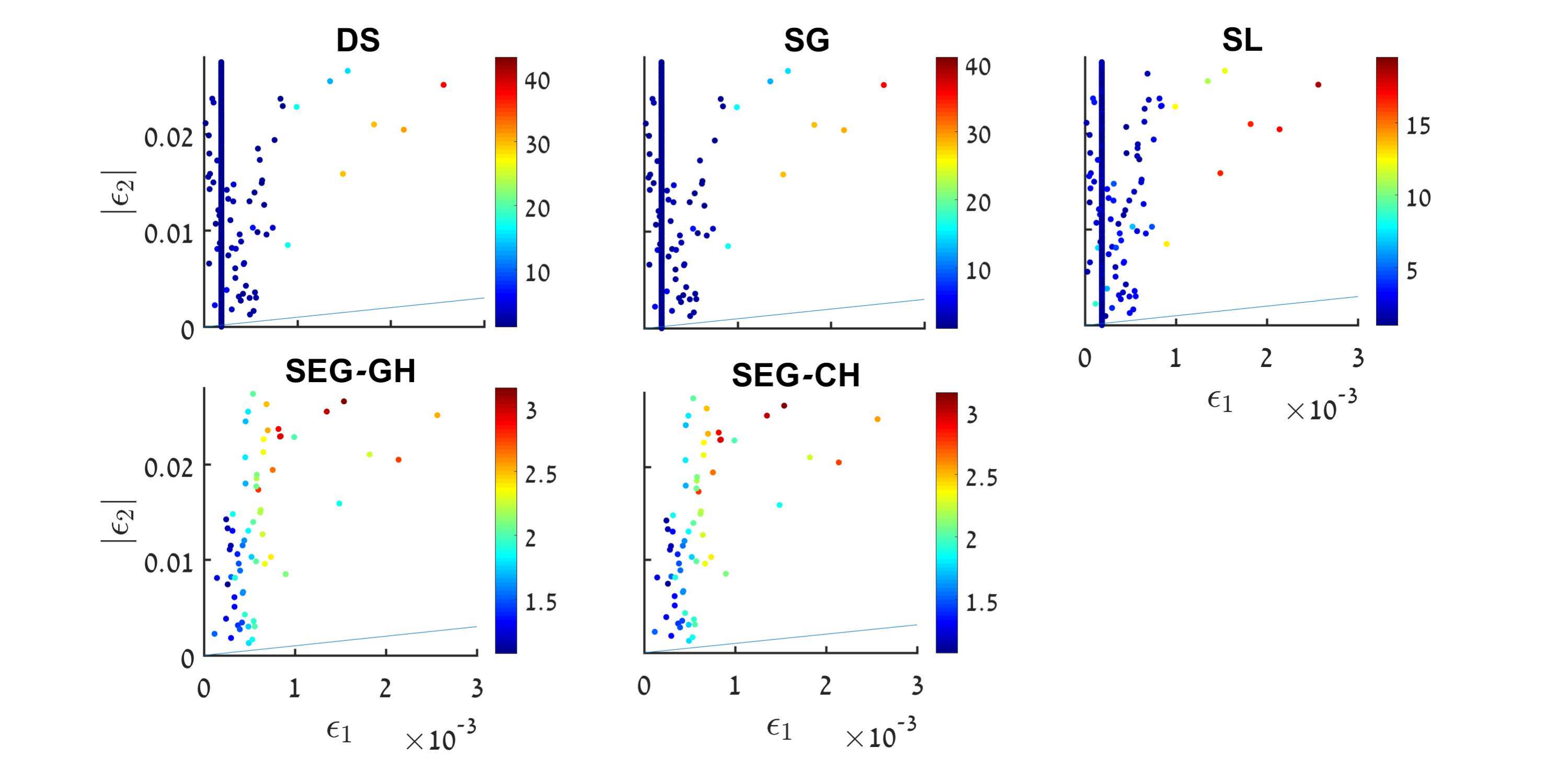}
\caption{Different analytical expressions and their errors relative to the exact numerical analysis, presented on the $\epsilon_1 - |\epsilon_2|$ plane. Each data point is the relative error between the analytic expression and the numerical result, and the color bars to the right of each panel indicate the percentage of relative error. The errors are filtered to show only errors above $1\%$, with numerical results $0.96<n_s<0.99$.}
\label{different_ns_METHODS2}
\end{figure}
\subsection{Possible explanations of the source of deviation between precise results and analytical estimates.}
From the discussion in Appendix B , one can easily see that the definition of $\nu$, is potentially the most significant discrepancy. The effect of this change in definition is an error of less than about $0.4\%$.

Table \ref{Examples} contains three examples of potentials. Two yield observables that are within acceptable limits, and a third shows an excluded precise result with an allowed analytic prediction. These examples are used to study the origin of discrepancy.
\begin{table}
	\begin{center}
		\begin{tabular}{|c||c|c|c|c|c|c|}
		\hline
		Ex. no.& $a_{1}$&$a_{2}$&$a_{3}$&$a_{4}$&$a_{5}$&$n_{s}$\\
		\hline
		1&$-0.01118$&$-0.0008$&$-0.2468$&$0.8726$&$-0.7825$&$0.9698$\\
		\hline
		2&$-0.01118$&$-0.0057$&$-0.2344$&$0.8631$&$-0.7804$&$0.9495$\\
		\hline
		3&$-0.01118$&$-0.0025$&$-0.1782$&$0.7100$&$-0.6916$&$0.9661$\\
		\hline
		\end{tabular}\caption{Shown are three examples for a 5 degree polynomial inflationary potentials. Examples no. 1 and 3 yield a precise result for $n_s$ which is well within the 68\% probability region. Example no.2 is the opposite case, with an analytic prediction within the 68\% region, but a precise result which is excluded. Tables \ref{Table_SlowVsPot} and \ref{ns of different methods}, refer to these potential examples. \label{Examples} }
		\end{center}
\end{table}
The differences between the slow-roll parameters defined via the potential vs. their definition in terms of time derivatives are also discussed in appendix B.
We have found, that in the  degree 5 polynomial potentials that were studied, small but significant departures from the relations in Eq. (B:\ref{approximated slow roll}) are detected.
For instance $\delta_{H}=-0.0016$ and $\delta_{V}=0.001$ at the time when $n_{s}$ is evaluated. Table \ref{Table_SlowVsPot} contains values of the three quantities $\epsilon_{H},\delta_{H},\tfrac{\delta_{H}\dddot{\phi}}{H\ddot{\phi}}$ as precisely calculated and analytically approximated, for three potentials of the 5 degree polynomial class. Table \ref{ns of different methods} contains the scalar index for the corresponding potentials (examples 1,2 and 3).
\begin{figure}[!ht]
\includegraphics[width=0.95\textwidth]{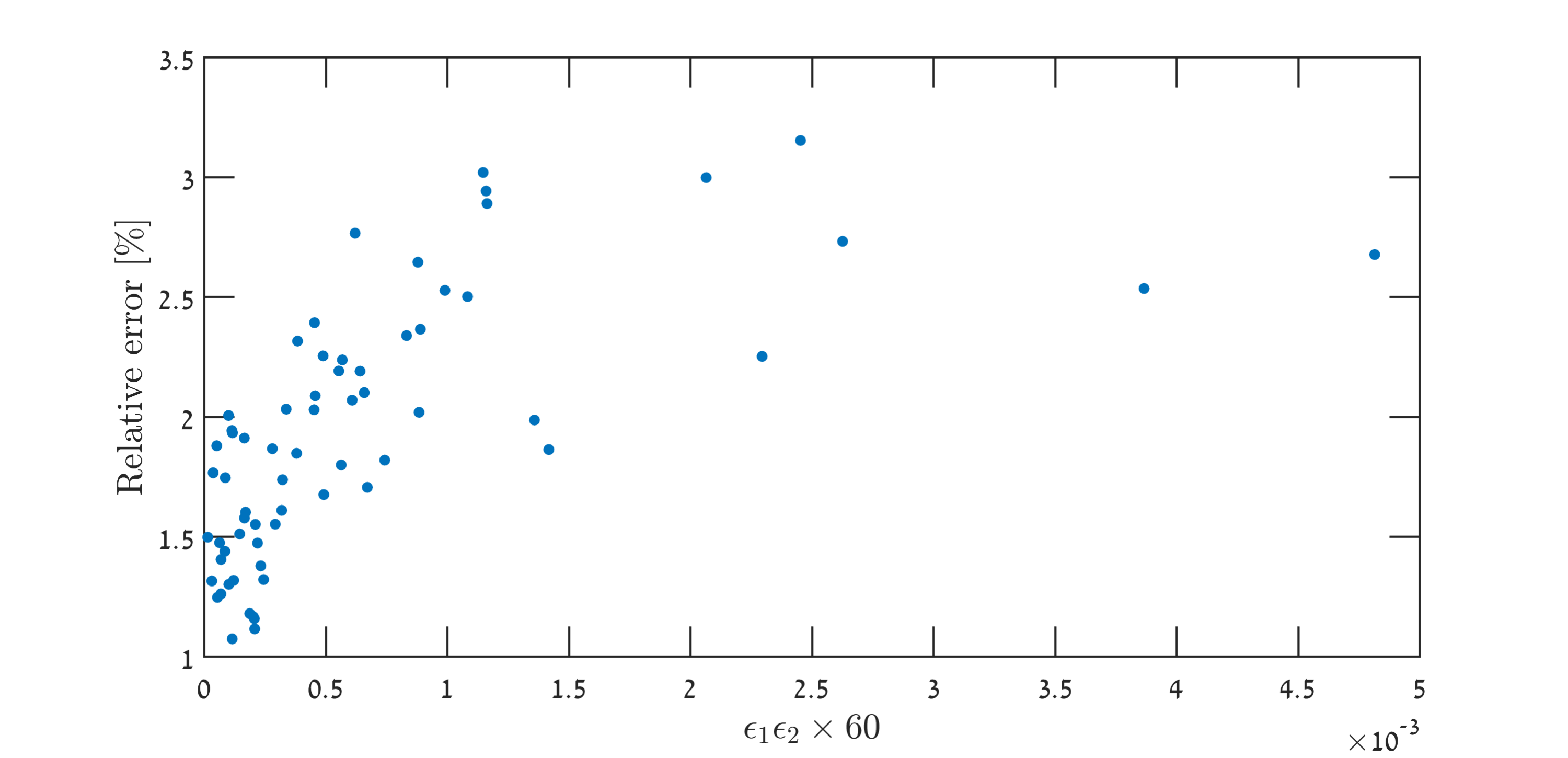}
\caption{While satisfying the condition $\epsilon_1 |\epsilon_2| \times \Delta N <10^{-2\sim 3}$, for $\Delta N = 60$, one finds a relative difference of well over $1\%$ between analytical predictions and numerical results. This is in contrast to the analysis proposed in \cite{Schwarz:2001vv}.}
\label{e1e260}
\end{figure}

\begin{table}[!h]
	\begin{center}
	\begin{tabular}{|l||c|c|c|}
		\hline
		Ex. no. & Quantity & slow roll value& pot. der. value\\
		\hline \hline
		 & $\epsilon$ & $6.28\cdot 10^{-5}$ & $6.24\cdot 10^{-5}$\\
		 1& $\delta$ & $-0.0068$ & $-0.0038$ \\
		 & $\frac{\delta \dddot{\phi}}{H\ddot{\phi}}$ &$0.0255$ & $0.0165$ \\
		 \hline
		  & $\epsilon$ & $6.23\cdot 10^{-5}$ & $6.20\cdot 10^{-5}$\\
		 2& $\delta$ & $0.0037$ & $0.0063$ \\
		 & $\frac{\delta \dddot{\phi}}{H\ddot{\phi}}$ &$0.0237$ & $0.0159$ \\
		 \hline
		  & $\epsilon$ & $6.26\cdot 10^{-5}$ & $6.23\cdot 10^{-5}$\\
		 3& $\delta$ & $-0.0016$ & $0.001$ \\
		 & $\frac{\delta \dddot{\phi}}{H\ddot{\phi}}$ &$0.0198$ & $0.0119$ \\
		 \hline
	\end{tabular}\caption{A table containing the three leading slow-roll parameters, as precisely calculated, vs. the values evaluated by the analytic approximation in Eq. (B:\ref{approximated slow roll}). While the difference in value for $\epsilon_H$ is negligible, the difference in $\delta_H$ might already be substantial and the difference for $\frac{\delta_{H}\dddot{\phi}}{H\ddot{\phi}}$ is significant.\label{Table_SlowVsPot}}
	\end{center}
\end{table}
The overall effect of this discrepancy can sometimes amount to a $5\sim 8\%$ error towards higher values.

Finally there is also a significant difference in the derivatives of $\nu$ and $\nu_{\text{\tiny{SL}}}$, $\nu_{\text{\tiny{SL}}}$ being $\nu$ in the SL formulation:
\begin{align}
	\nu_{\text{\tiny{SL}}}=\frac{3+2\delta_H +\epsilon_H}{2(1-\epsilon_H)}\;,
\end{align}
where time dependency of the slow roll parameters is neglected. This difference is mainly due to neglecting the term $\frac{\delta_{H}\dddot{\phi}}{H\ddot{\phi}}$ in the definition of $\frac{Z''}{Z}$. This yields a difference in the derivative terms of the order of $0.02\sim 0.04$, which in turn is responsible for a difference in $n_{s}$ of the order of $4\sim 8\%$. Using $\nu_{\text{\tiny{SL}}}$ instead of the full term, tends to drive the resulting $n_s$ downwards.

The tendencies of the two aforementioned errors are opposite, and so they might sometimes cancel each other. This makes it possible to get an accurate result using the standard SL expression for a specific potential, but studying a collection of such potentials reveals the incomplete nature of this cancellation.

Table \ref{ns of different methods} shows the different results using different methods of deriving the scalar index. We use three different analytical methods: (1) Eq. (B:\ref{SLR}) - The SL original method, extracting a term for the scalar index as a function of the potential and its derivatives, (2) Eq. \eqref{nsH} - The SL original method, but not relating slow roll quantities to potential and derivatives, and (3) Using the same methods as the SL analysis, with the definition for $\nu$ as in Eqs. (B:\ref{TrueNu},\ref{AccurateZoZ-2},\ref{TrueNs}). From this analysis it seems the origin of the most significant error is the inaccurate relations between slow roll parameters and their potential and derivatives counterparts. Second in significance is the definition of $\nu$ with the full $\frac{Z''}{Z}\tau^{2}$ expression, along with the proper derivation of $\frac{\partial \nu}{\partial \log(k)}$.  The evaluation of $-\tau a H (1-\epsilon_{H})=1$ is off by $\sim 0.04\%$ and the difference between $\psi(\tfrac{3}{2})$ and $\psi(\nu)$  yields a correction of the order of $\sim 0.01\%$.

There might be additional factors that stem from the temporal dependence of $\nu$ in the MS equation, however, these mostly affect the running of the spectral index, and are harder to estimate accurately.

Taking these approximations into account, lowers the discrepancy to the order of $0.5\%$, in a consistent manner.
Another possible explanation is that the time-dependence in \eqref{Z''/Z}, modifies the corresponding $\omega^{2}_{k}(\tau)=\left(k^{2}-\frac{\tilde{C}}{\tau^{2}}\right)$ to $\omega^{2}_{k}(\tau)=\left(k^{2}-\frac{f(\tau)}{\tau^{2}}\right)$.
This could lead to modified solutions for the MS equation. An example of this phenomenon is given in \cite{Martin:2000ei}, where the Hankel functions were replaced by the Whittaker functions (albeit these models are observationally excluded). It is worth mentioning that this avenue was studied analytically by Dodelson \& Stewart \cite{Stewart:2001cd,Dodelson:2001sh}. They derived an expression for the scalar index in cases where the slow-roll hierarchy breaks down. However, this analysis was not checked numerically. Additional derivation attempts aiming at yielding better precision analytical expression for the scalar index $n_s$ were made in \cite{Gong:2001he,Schwarz:2001vv}. Specifically \cite{Schwarz:2001vv} supplies an analysis of the predicted level of accuracy as a function of the horizon flow functions $\epsilon_1\equiv \epsilon_H$ and $\epsilon_2\equiv 2(\epsilon_H +\delta_h)$, in figure \ref{Different_regions}. The different approximation schemes were put to the numerical test in the context of our models. Figure \ref{different_ns_SEG} shows that all methods of approximation yield results varying in accuracy and precision levels, it also shows however that the SEG approximation is the best candidate to improve on, since on average they yield errors of less than $1\%$.
~ Studying results where relative errors in $n_s$ are over $1\%$, for each expression and locating it on the $\epsilon_1-|\epsilon_2|$ diagram in figure \ref{different_ns_METHODS2} reveals that the analysis offered in \cite{Schwarz:2001vv} is not completely applicable to our models. Figure \ref{e1e260} shows that the for the models studied, even though the conditions outlined in \cite{Schwarz:2001vv} are met, and $\epsilon_1 \epsilon_2 \times \Delta N <10^{-2\sim 3}$ for $\Delta N=60$, the relative error between numerical result and SEG-CH expression can be above $1\%$.  

\section{Summary, conclusions and outlook}
An interesting class of models that can produce a high tensor-to-scalar ratio while conforming to observable values of $n_{s}$ and $n_{\text{\tiny{run}}}$ was presented and studied. This work  has shown that while the arguments for small field model validity presented in [\cite{BenDayan:2009kv,Hotchkiss:2011gz} generally apply, the method by which they choose favoured models is based on  approximations that are not always accurate enough for the cases studied. While this work argued this possible weakness, it also supplied a remedy: The precise calculation method. Using precise calculations points to new candidates previously disregarded.
\begin{table}
		\begin{center}
		\begin{tabular}{|c|c||c|c|c|c|}
			\hline
				Ex. no. & & Num. value &Eq. (B:\ref{SLR}) &Eq. \eqref{nsH}& Eq. (B:\ref{TrueNu},\ref{AccurateZoZ-2},\ref{TrueNs})\\
			\hline		
			1 &$n_{s}$ & $0.9698$ & $0.9833$ & $1.05$ & $0.9650$\\
			  &rel. error& $0$/NA & $1.38\%$& $7.99\%$& $-0.49\%$\\
			\hline		
			2 &$n_{s}$ & $0.9495$ & $0.9643$ & $1.027$ & $0.9474$\\
			  &rel. error& $0$/NA & $1.54\%$& $7.8\%$& $-0.21\%$\\
			\hline
			3&$n_{s}$ &$ 0.9661 $&$ 0.9803 $&$ 1.031 $&$ 0.9695$\\
			&rel. error& $0$/NA & $1.4\%$& $6.6\%$& $0.35\%$\\
			\hline
		\end{tabular}
	\caption{Shown are different results for different methods of calculating the scalar index $n_s$. These were calculated for the 3 example potentials mentioned in Table \ref{Examples}. The first is the numerical result. Next is the standard Stewart \& Lyth expression Eq.(B:\ref{SLR}). Another result is given by using \eqref{nsH}, without substituting potential and derivative expressions for slow roll parameters. Finally we use Eqs.(B:\ref{AccurateZoZ-2},\ref{TrueNu},\ref{TrueNs}), to accurately assess the scalar index.\label{ns of different methods}}
	\end{center}
\end{table}
Specifically, The predictions made using the standard SL analytic expressions were found to deviate by more than $1\%$ from the actual results, for many models in this class. Other approximate expressions such as those suggested in \cite{Dodelson:2001sh,Gong:2001he,Schwarz:2001vv} are, in general, better than the SL expressions, but still miss by more than $1\%$ in some cases.

We hope to extend this work  to models that produce higher values of r, and determine the best candidate for small field inflationary models \cite{Wolfson:2018lel}.

\section*{Acknowledgements}
We would like to thank Ido Ben-Dayan for stimulating conversations and insights regarding the inflationary models. Rahul Kumar was also instrumental in helping develop the underlying numerical package. Additional helpful discussions and remarks were given by Antonio Riotto, as well as Misao Sasaki and Smadar Naoz. \\
\bigskip The research was supported by the Israel Science Foundation grant no. 1294/16.

\section*{Supporting information}

\section*{Appendix A - Benchmark tests}

\begin{figure}[!h]
\includegraphics[width=1\textwidth]{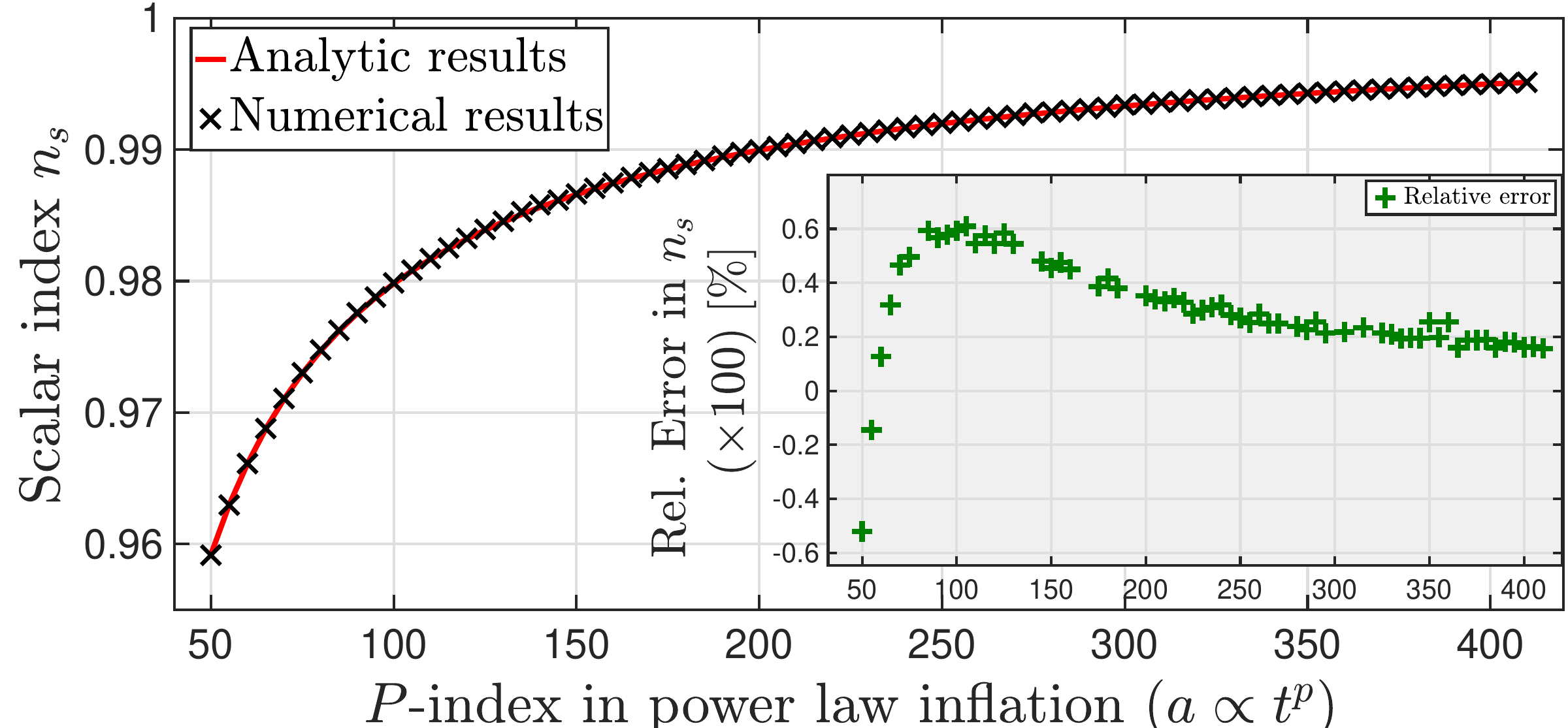}
\caption{Shown is the convergence of results of precise calculations (crosses) to analytic ones (solid curve). As the power law index grows, convergence to a de-Sitter inflation is apparent. The embedded panel demonstrates the high accuracy level, as the initial typical error is of order $10^{-2}\sim 10^{-3}\%$ and it decreases as a function of the power law index $p$. \label{fig:convergence+dS}}
\end{figure}
\subsubsection{Power law inflation}\label{sec:power law inflation}

The accuracy of the procedure is tested by using the benchmark case of power law inflation ($a\propto t^{p}$). This is the only case for which the analytic results are exact since $\epsilon_H$  and $\delta_H$ are constants $\epsilon_H=\frac{1}{p}$, and $\delta_H=-\frac{1}{p}$. The cosmological parameters are given by:
\begin{align}
\left\{\begin{array}{c}
n_{s}=1-\frac{2}{p}\;,\\
n_{\text{\tiny{run}}}= 0 .
\end{array}\right.
\end{align}
\begin{figure}[!h]
\includegraphics[width=1\textwidth]{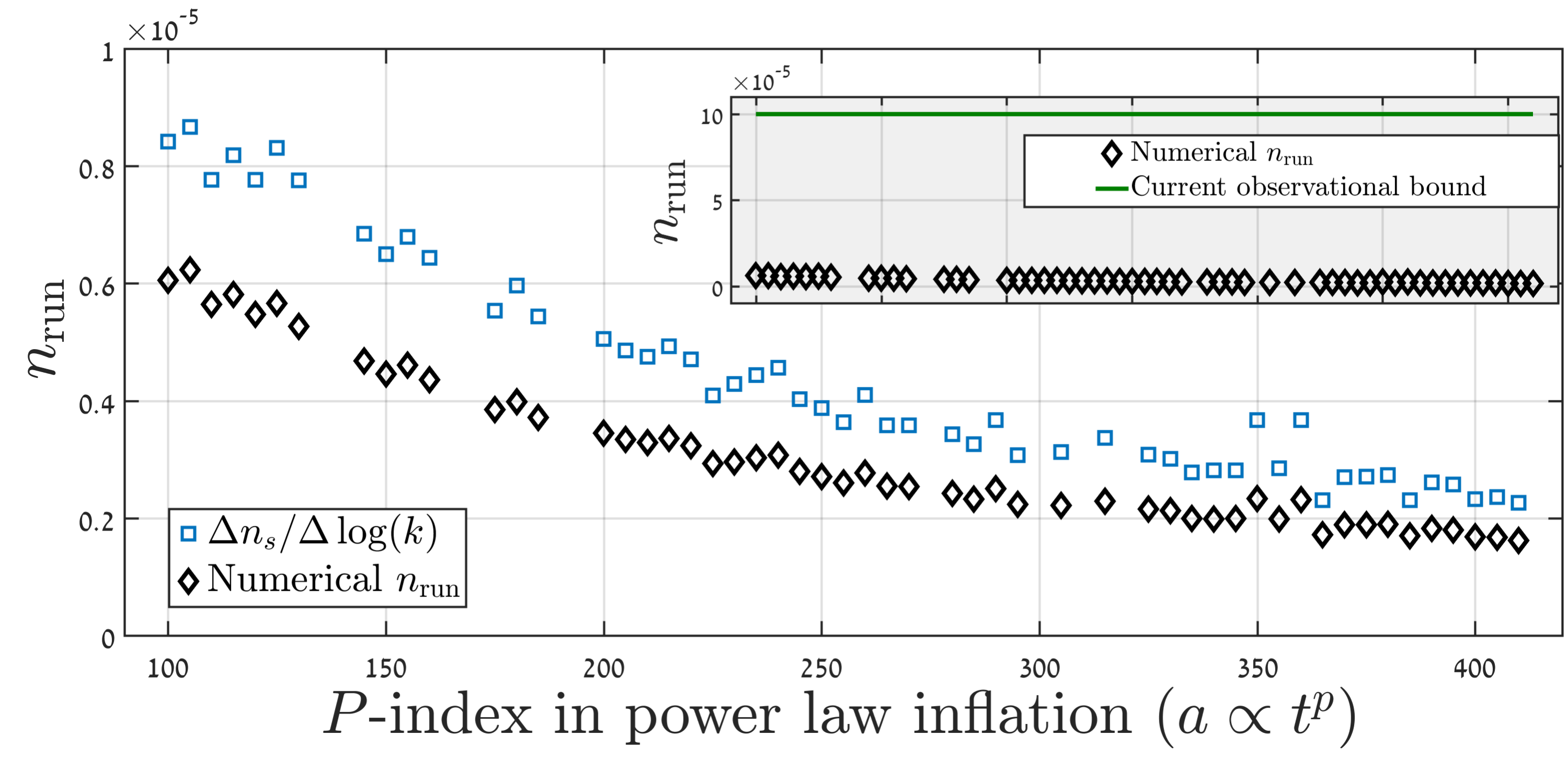}
\caption{Convergence of the numerical results for $n_{\text{\tiny{run}}}$ to the analytic value. Shown are the numerical results for $n_{\text{\tiny{run}}}$ (diamonds) and the diagnostic $\frac{\Delta n_{s}}{\Delta \log(k)}$ (squares). Also, we show that the values for $n_{\text{\tiny{run}}}$ as extracted, are well below current observational bound. As such the accuracy levels in $n_{\text{\tiny{run}}}$ are sufficient. \label{fig:nrunAccuracy}}
\end{figure}
In Fig.~\ref{fig:convergence+dS}, results of the convergence of the precise calculations to analytic predictions for the case of power law inflation are shown.
The overall shape of both precise calculations and analytic curves agree and a relative error in $n_{s}$,\\
estimated by:
$
\left(n_{s}^{\text{\tiny{precise}}}-n_{s}^{\text{\tiny{analytic}}}\right)/ \frac{1}{2} \left(n_{s}^{\text{\tiny{precise}}}+n_{s}^{\text{\tiny{analytic}}}\right)\;,
$
is of the order of $10^{-2}\sim 10^{-3}\%$. The method for error estimate in $n_{\text{\tiny{run}}}$ is more subtle, since the correct value of $n_{\text{\tiny{run}}}$ is zero. In order to assess our error in $n_{\text{\tiny{run}}}$ the following diagnostic was therefore used: the difference between the precise and analytic $n_{s}$ is divided by the  difference in $\log(k)$. The criterion for convergence is that the absolute value of $n_{\text{\tiny{run}}}$ is smaller than $\frac{\Delta n_{s}}{\Delta \log(k)}$.

Figure \ref{fig:nrunAccuracy} displays the convergence of the precisely calculated results for $n_{\text{\tiny{run}}}$, using the diagnostic $\frac{\Delta n_{s}}{\Delta \log(k)}$. It is apparent that $n_{\text{\tiny{run}}}$ is always bounded from above by $\frac{\Delta n_{s}}{\Delta \log(k)}$. Additionally the extracted $n_{\text{\tiny{run}}}$ is an order of magnitude or so below current observational bound.

\subsubsection{Quadratic potentials}
\begin{figure}[!h]
\includegraphics[width=1\textwidth]{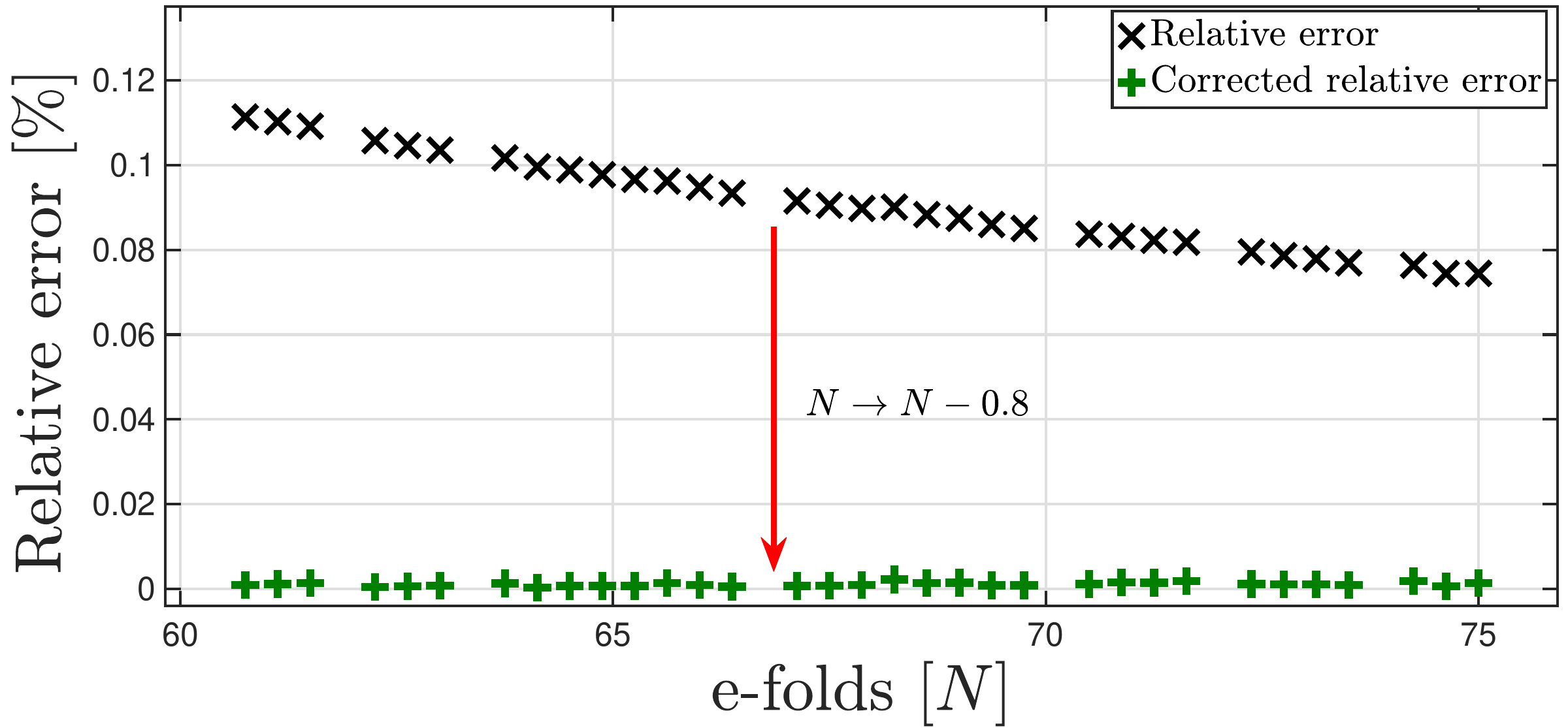}
\caption{Relative error (in percents) between numerical results and the SL analytical expression (black X's). The errors converge to $0$ for large values of $N$. Shifting the number of the efolds by $N\rightarrow N-0.8$ yields a relative error of the order of $10^{-3} \sim 10^{-4} \%$ (green pluses).  \label{quad1} }
\end{figure}
As we aim to study models that produce slow roll parameters which are time dependent, we need to check the precision of the numerical code against such models.

Consequently we tested the accuracy of our calculations for quadratic potentials of the type
\begin{align}
V=\frac{1}{2}m^{2}\phi^{2}.
\end{align}
In these cases the analytic expression for the scalar index is given by,
\begin{align}
n_{s}=1-\frac{8}{4N+2}+\frac{32b}{\left(4N+2\right)^2},\label{SL-quad}
\end{align}
Here $N$ is the number of efolds and $b$ is the same as in \eqref{nsH}.
Figure~\ref{quad1} presents the results of this study, as relative errors between precise calculations and the SL analytic expressions. These results are accurate to $\sim 0.1 \%$. However there is a systematic error that is traced back to the inaccuracy of the approximation:
\begin{align}
	N=\int_{t_{\text{\tiny{CMB}}}}^{t_{\text{\tiny{end}}}}Hdt \simeq -\int_{\phi_{\text{\tiny{CMB}}}}^{\phi_{\text{\tiny{end}}}}\frac{V}{V'}d\phi.
\end{align}
A shift $N\rightarrow N-0.8$ is sufficient to reduce the systematic error such that the relative error is of the order of $10^{-3}\sim 10^{-4}\%$. Additional types of simple potentials, which yield time-dependent slow-roll parameters were also studied. In all cases the relative error between calculated results and the traditional SL expression \eqref{nsV} is bounded from above by $\sim 0.1\%$. Furthermore, a more careful analytical treatment leads to better accuracy, bounded from above by about 0.02\% relative error. Additionally, we were able to recover the ``Cosmic ring'' phenomenon, that is the PPS response to a step function in the potential. This response feature in the PPS was first studied in \cite{Adams:2001vc}.

We take all these results as a strong indication of sufficient accuracy of our calculations.

\section*{Appendix B - A short recap of the Stewart-Lyth formulation}\label{AppendixB}
In order to better understand the origin of discrepancy between precise results and the analytical SL expression, we retrace the procedure of deriving an analytical expression for $n_s$. Recalling the definition for the pump field $Z$, and the MS equation (Eqs. \eqref{Z},\eqref{MSfirst}). The parameter $\nu$ is properly defined as:
\begin{align}
	\nu=+\sqrt{\frac{Z''}{Z}\tau^{2}+\frac{1}{4}}.\label{TrueNu}
\end{align}
However, in the SL formulation, the approximations made lead to the defining of $\nu$ as:
\begin{align}
	\nu_{\text{\tiny{SL}}}=\frac{3+2\delta_H +\epsilon_H}{2\left(1-\epsilon_H\right)},
\end{align}
which can be very different.
Then,
\begin{align}
	U_{K}'' +\left(k^{2}-\frac{\left(\nu^{2}-\frac{1}{4}\right)}{\tau^{2}}\right)U_{k}=0.
\end{align}
For a constant $\nu$ this becomes the Bessel equation, with known solutions.
As mentioned before \eqref{Z''/Z}, the value of $\tfrac{Z''}{Z}$ is given by:
\begin{align}
	\frac{Z''}{Z}=2a^{2}H^{2}\left(1+\frac{3\delta_{H}}{2}+\epsilon_{H} +\frac{\delta_{H}^{2}}{2}+\frac{\epsilon_{H}\delta_{H}}{2} +\frac{1}{2H}\left(\dot{\epsilon_{H}}+\dot{\delta_{H}}\right)\right).\label{AccurateZoZ}
\end{align}
In many cases, one assumes that the time derivatives are small and can be neglected. However, these derivatives yield 2nd order terms that can significantly affect the value of $\tfrac{Z''}{Z}$. The full expression is given by:
\begin{align}
	\frac{Z''}{Z}=2a^{2}H^{2}\left(1+\frac{3\delta_{H}}{2} +\epsilon_{H}+\epsilon_{H}	^{2} +2\epsilon_{H}\delta_{H} +\frac{1}{2}\frac{\delta_{H}\dddot{\phi}}{H\ddot{\phi}}\right), \label{AccurateZoZ-2}
\end{align}
which may  differ from Eq.~(\ref{AccurateZoZ}) when $\delta^2_{H}$ and/or $\frac{\delta_{H}\ddot{\phi}}{H\dot{\phi}}$ are non-negligible. $\epsilon^2_H$ is usually of the order of $10^{-5}$ or less, even for models with high $r$.

Applying boundary conditions and taking the small arguments limit we are left with a power spectrum of:
\begin{align}
	\log\left(P_{R}\right)&=-\log(32\pi^{2}\Gamma^{2}(\tfrac{3}{2}))&   \\ \nonumber  &+ 2\nu \log(2) +2 \log(k) +2 \log\left(\Gamma(\nu)\right) \\ \nonumber &+(1-2\nu) \log(-k\tau),&
\end{align}
which yields the scalar index of:
\begin{align}
	n_{s}=4-2\nu +2\left(\log(2) +\psi(\nu)\right)\frac{\partial \nu}{\partial \log(k)},\label{TrueNs}
\end{align}
with the digamma function $\psi(x)\equiv \frac{\Gamma'(x)}{\Gamma(x)}$.
The final expression is heavily dependent on the value and time derivative of $\nu$. This is a possible source of discrepancy. 
It is now customary to define:
\begin{align}
	\begin{array}{ccc}
	\alpha=\left(\frac{V_{,\phi}}{V}\right)^{2}& \beta =\frac{V_{,\phi\phi}}{V} & \gamma=\frac{V_{,\phi^{3}}}{V_{,\phi}},\\
	\end{array}
\end{align}
or related quantities ($\varepsilon_{V}=\frac{\alpha}{2}$ for instance).
Having defined these, usually one connects the original slow roll parameters with the above quantities by \cite{Stewart:1993bc}
\begin{align}
	\begin{array}{c}
		\epsilon	_{H}\simeq \frac{\alpha}{2}-\frac{\alpha^{2}}{3}+\frac{\alpha\beta}{3}\\
		\delta_{H}\simeq \frac{\alpha}{2} -\beta -\frac{2\alpha^{2}}{3}+\frac{4\alpha\beta}{3} -\frac{\beta^{2}}{3}-\frac{\alpha\gamma}{3}\\
		\frac{\delta_{H}\dddot{\phi}}{H\ddot{\phi}}\simeq \alpha^{2}-\frac{5\alpha\beta}{2}+\beta^{2}+\alpha\gamma.
	\end{array}\label{approximated slow roll}
\end{align}
With these relations one can substitute the slow-roll parameters in Eq. \eqref{nsH}, for the quantities in Eq. \eqref{approximated slow roll}, to get the most commonly used analytical expression for the scalar index \cite{Lyth:1998xn}:
\begin{align}
	\nonumber n_{s}&\simeq 1-6\varepsilon_{V}+2\eta_{V} &\\ \nonumber&+2\times\bigg[\frac{\eta_{V}^2}{3}-(8b+1)\varepsilon_{V}\eta_{V}&& \\&-\left(\frac{5}{3}-12b\right)\varepsilon_{V}^2+\left(b+\frac{1}{3}\right)\xi_{V}^2\bigg],& \label{SLR}
\end{align}
where $\varepsilon_{V}=\tfrac{\alpha}{2}\;;\;\eta_{V}=\beta\;;\;\xi^2_{V}=\alpha\gamma$, and with the same $b$ as in \eqref{nsH}.


\end{document}